\documentclass[12pt,a4paper,aps,preprint,superscriptaddress,nofootinbib]{revtex4-1}
\usepackage[utf8]{inputenc}
\usepackage{graphicx}
\usepackage{amssymb}
\usepackage{textcomp}
\usepackage{amsmath}
\usepackage{bm}
\usepackage{times}
\usepackage{color}
\usepackage{multirow}
\usepackage[colorlinks=true, pdfstartview=FitV, linkcolor=blue, citecolor=blue, urlcolor=blue]{hyperref}
\allowdisplaybreaks[4]

\def\lsim{\mathrel{\raise.3ex\hbox{$<$\kern-.75em\lower1ex\hbox{$\sim$}}}}
\def\gsim{\mathrel{\raise.3ex\hbox{$>$\kern-.75em\lower1ex\hbox{$\sim$}}}}

\newcommand{\calO}{{\mathcal{O}}}
\newcommand{\calB}{{\mathcal{B}}}

\newcommand{\GeV}{{\rm GeV}}
\newcommand{\TeV}{{\rm TeV}}
\newcommand{\Tr}{{\rm Tr}}

\definecolor{orange}{rgb}{1,0.5,0}

\begin{document}

\title{Implication of $K\to \pi \nu \bar{\nu}$ for generic neutrino interactions in effective field theories}

\author{Tong Li}
\email{litong@nankai.edu.cn}
\affiliation{
School of Physics, Nankai University, Tianjin 300071, China
}
\author{Xiao-Dong Ma}
\email{maxid@phys.ntu.edu.tw}
\affiliation{
Department of Physics, National Taiwan University, Taipei 10617, Taiwan
}
\author{Michael A. Schmidt}
\email{m.schmidt@unsw.edu.au}
\affiliation{
School of Physics, The University of New South Wales, Sydney, New South Wales 2052, Australia
}

\begin{abstract}
In this work we investigate the implication of $K\to \pi \nu \bar{\nu}$ from
the recent KOTO and NA62 measurements for generic neutrino interactions and the new
physics scale in effective field theories. The interactions between quarks and
left-handed Standard Model (SM) neutrinos are first described by the low energy effective field theory (LEFT) below the electroweak
scale. We match them to the chiral perturbation theory ($\chi$PT) at the chiral symmetry breaking scale to
calculate the branching fractions of Kaon semi-invisible decays and match them
up to the SM effective field theory (SMEFT) to constrain new physics above the electroweak scale. In the
framework of effective field theories, we prove that the Grossman-Nir bound is
valid for both dim-6 and dim-7 LEFT operators, and the dim-6 vector and scalar operators
dominantly contribute to Kaon semi-invisible decays based on LEFT and chiral power counting rules. They are induced by multiple dim-6 lepton-number-conserving operators and one dim-7 lepton-number-violating operator in the SMEFT,
respectively. In the lepton-number-conserving $s\to d$ transition, the $K\to \pi \nu \bar{\nu}$ decays provide the most
sensitive probe for the operators with $\tau\tau$ component and point to a corresponding new physics scale of $\Lambda_{\rm NP}
\in[47~\text{TeV},~72~\text{TeV}]$ associated with a single effective coefficient.
The lepton-number-violating operator can also explain the observed
$K\to\pi\nu\bar{\nu}$ discrepancy with the SM prediction within a narrow range $\Lambda_{\rm NP}\in
[19.4~\text{TeV},~21.5~\text{TeV}]$, which is consistent with constraints from Kaon invisible decays.
\end{abstract}

\maketitle

\section{Introduction}
\label{sec:Intro}

Recently, the KOTO experiment at J-PARC~\cite{koto201618-1,koto201618-2} and the NA62 experiment at CERN~\cite{NA62} announced preliminary results of Kaon semi-invisible decays~\cite{Kitahara:2019lws}
\begin{eqnarray}\label{koto}
\calB_{K_L\to \pi^0\nu \bar{\nu}}^{\text{KOTO16/18}}&=&2.1^{+4.1}_{-1.7}\times 10^{-9},
\\\label{na62}
 \calB_{K^+\to \pi^+ \nu \bar{\nu}}^{\text{NA62}}&<&2.44\times 10^{-10} ,
\end{eqnarray}
at the 95\% confidence level (CL). They update the upper limit on the decay rate of $K^+\to \pi^+\nu\bar{\nu}$ from BNL E949~\cite{Artamonov:2008qb,Artamonov:2009sz} and the limit on the branching ratio $\calB(K_L\to \pi^0\nu \bar{\nu})$ from the 2015 run at KOTO itself~\cite{Ahn:2018mvc}
\begin{eqnarray}
\calB_{K_L\to \pi^0\nu \bar{\nu}}^{\text{KOTO15}}&<&3.0\times 10^{-9},
\\
\calB_{K^+\to \pi^+ \nu \bar{\nu}}^{E949}&<&3.35\times 10^{-10},
\end{eqnarray}
at the 90\% CL.
These decays are mediated by flavor changing neutral currents (FCNC) and thus are suppressed by
the GIM mechanism in the Standard Model (SM), giving $\calB_{K_L\to \pi^0\nu \bar{\nu}}^{\text{SM}}=(3.4\pm 0.6)\times 10^{-11}$ and $\calB_{K^+\to \pi^+ \nu \bar{\nu}}^{\text{SM}}=(8.4\pm 1.0)\times 10^{-11}$, respectively~\cite{Buras:2006gb,Brod:2010hi,Buras:2015qea}. In the SM no events are expected from the above Kaon rare decays, but KOTO reported three signal events in the search of $K_L\to \pi^0 \nu\bar{\nu}$.
There exist quite a few works trying to explain these intriguing events reported by KOTO~\cite{Kitahara:2019lws,Fabbrichesi:2019bmo,Egana-Ugrinovic:2019wzj,Dev:2019hho} (or constrain particular new physics (NP) models~\cite{Fuyuto:2014cya,Mandal:2019gff,Calibbi:2019lvs}) and meanwhile avoid the violation of its relation with the $K^+\to \pi^+ \nu\bar{\nu}$ decay, that is the Grossman-Nir bound~\cite{Grossman:1997sk}. These efforts require the introduction of a new invisible degree of freedom with the mass scale being around $100-200$ MeV.

The interpretation of the KOTO result depends on not only whether the invisible particles are viewed as neutrinos, but also the experimental uncertainties. Even if we only take into account the statistical uncertainties at 95\% CL for neutrino final states, there is allowed space for heavy NP beyond the SM consistent with both $K_L\to \pi^0 \nu\bar{\nu}$ and $K^+\to \pi^+ \nu\bar{\nu}$ measurements and satisfying the Grossman-Nir bound. As one can see from the Fig.~1 in Ref.~\cite{Kitahara:2019lws}, the allowed region is rather delimited and not far away from the SM prediction. It can provide a constraint on the relevant quark-neutrino interactions and shed light on the search for generic neutrino interactions in the future. Thus, without introducing any new light particles, we focus on heavy NP contributing to the generic quark-neutrino interactions and generically confine the NP scale from the allowed region of $\calB(K_L\to \pi^0\nu \bar{\nu})$ and $\calB(K^+\to \pi^+ \nu \bar{\nu})$ measurements.
As the neutrino flavor is not measured and the fermionic nature of neutrinos is not determined, the semi-invisible Kaon decays $K\to \pi \nu \bar{\nu}$ are sensitive probes for a range of interactions.

In this work, we will use an effective field theory approach, where NP is described
by a set of non-renormalizable operators which are added to the SM Lagrangian
\begin{eqnarray}
\mathcal{L}_{\rm eff}= \mathcal{L}_{\rm SM} + \sum_i \sum_{d\geq 5} C_i^{(d)} \calO_i^{(d)} \; .
\end{eqnarray}
Here $\calO_i^{(d)}$ are the dimension-$d$ (dim-$d$ in short below) effective operators. Each Wilson coefficient $C_i^{(d)}$ is associated with a NP scale $\Lambda_{\rm NP}=(C_i^{(d)})^{1/(4-d)}$.
We first use the low energy effective field theory (LEFT)~\cite{Jenkins:2017jig,Jenkins:2017dyc} to
describe the interactions between quarks and left-handed SM
neutrinos below the electroweak scale. Then, in order to calculate the
Kaon decay rate, we match the LEFT operators to chiral
perturbation theory ($\chi$PT)~\cite{Gasser:1983yg,Gasser:1984gg} at the chiral symmetry breaking scale
to take into account non-perturbative QCD effects. The branching
fractions of Kaon semi-invisible decays are evaluated in terms of
the Wilson coefficients and neutrino bilinears as external sources.
Finally, we match them up to the Standard Model effective field
theory (SMEFT) to constrain new physics above the electroweak
scale~\cite{Buchmuller:1985jz,Grzadkowski:2010es, Lehman:2014jma,Liao:2016hru,Henning:2015alf,Liao:2016qyd}.

The paper is outlined as follows. In Sec.~\ref{sec:LEFTandKpi}, we describe the LEFT basis and give the quark-neutrino operators relevant for our study.
The LEFT operators are matched to $\chi$PT and we show the general expressions for the branching fractions of Kaon semi-invisible decays. We then match the results to the SMEFT in Sec.~\ref{sec:SMEFT}. In Sec.~\ref{sec:Implication} we show the implication of $K\to \pi \nu \bar{\nu}$ for new physics and discuss other constraints. Our conclusions and some discussions are drawn in Sec.~\ref{sec:Con}.
Some calculation details for Kaon decays are collected in the Appendix.

\section{Generic neutrino interactions and $K\to \pi \nu\bar{\nu}$ calculation in $\chi$PT}
\label{sec:LEFTandKpi}

\subsection{Generic quark-neutrino operators in LEFT basis}
\label{sec:LEFT}

We consider the effective operators for neutrino bilinears coupled to SM quarks in the framework of LEFT obeying SU$(3)_{\rm c}\times$U$(1)_{\rm em}$  gauge symmetry. In the basis of LEFT for neutrinos, the only dim-5 operator contributing to the neutrino magnetic moments is~\cite{Canas:2015yoa}
\begin{eqnarray}\label{numm}
\calO_{\nu\nu F}&=&(\overline{\nu^C}i\sigma_{\mu\nu}\nu)F^{\mu\nu}+h.c. \; ,
\end{eqnarray}
where $F_{\mu\nu}$ is the electromagnetic field strength tensor
and $\nu =P_L \nu$ denote left-handed active SM neutrinos.
Its SMEFT completion has been investigated by Cirigliano et al.~in Ref.~\cite{Cirigliano:2017djv}.
In principle, the neutrino magnetic moment operator can yield the $K\to \pi \nu \nu$ process through a long-distance contribution with one vertex connecting to the $s\to d\gamma$ transition operator $\bar s\sigma_{\mu\nu}P_{L/R}dF^{\mu\nu}$. The corresponding coefficient is estimated to be $C_{sdF}\sim 10^{-9}~\GeV^{-1}$ in the SM~\cite{Tandean:1999mg}. There also exists a strong constraint on $|C_{\nu\nu F}|\leq 4\times10^{-9}~\GeV^{-1}$~\cite{Canas:2015yoa}. We thus conclude that the contribution from this operator to $K\to\pi\nu\nu$ transition is negligible.
There are also the dim-6 operators~\cite{Jenkins:2017jig} with lepton number conservation (LNC, $|\Delta L|=0$)
\begin{align}
\calO_{u\nu 1}^V&=(\overline{u_L}\gamma^\mu u_L)(\overline{\nu}\gamma^\mu \nu)\; , &
\calO_{d\nu 1}^V&=(\overline{d_L}\gamma^\mu d_L)(\overline{\nu}\gamma^\mu \nu)\; ,\\
\calO_{u\nu 2}^V&=(\overline{u_R}\gamma^\mu u_R)(\overline{\nu}\gamma^\mu \nu)\; ,&
\calO_{d\nu 2}^V&=(\overline{d_R}\gamma^\mu d_R)(\overline{\nu}\gamma^\mu \nu)\; ,
\end{align}
and those with lepton number violation (LNV, $|\Delta L|=2$)
\begin{align}
\calO_{u\nu 1}^S&= (\overline{u_R} u_L)(\overline{\nu^C} \nu)+h.c.\; ,&
\calO_{d\nu 1}^S&= (\overline{d_R} d_L)(\overline{\nu^C} \nu)+h.c.\; ,\\
\calO_{u\nu 2}^S&= (\overline{u_L} u_R)(\overline{\nu^C} \nu)+h.c.\; ,&
\calO_{d\nu 2}^S&= (\overline{d_L} d_R)(\overline{\nu^C} \nu)+h.c.\; ,\\
\calO_{u\nu}^T&= (\overline{u_R} \sigma^{\mu\nu}u_L)(\overline{\nu^C}\sigma_{\mu\nu} \nu)+h.c.\; ,\label{tensor}&
\calO_{d\nu}^T&= (\overline{d_R} \sigma^{\mu\nu}d_L)(\overline{\nu^C}\sigma_{\mu\nu} \nu)+h.c.\; .
\end{align}
where $u_L(u_R)$ and $d_L(d_R)$ denote the left- (right-) handed up-type and down-type quark fields in mass basis, respectively. Note that the tensor operator $\overline{\nu_\alpha^C} \sigma^{\mu\nu} \nu_\beta$ vanishes for identical neutrino flavors (with $\alpha=\beta$).
The flavors of the two quarks and those of the two neutrinos in the above operators can be different although we do not specify their flavor indexes here.
For the notation of the Wilson coefficients, we use the same subscripts as the operators, for instance $C_{d\nu1}^{V,xy\alpha\beta}$ together with $\calO_{d\nu1}^{V,xy\alpha\beta}$, where $x,y$ denote the down-type quark flavors and $\alpha,\beta$ are the neutrino flavors.
In the following we will study the $K\to \pi$ transition and thus only consider the operators with down-type quarks $s$ and $d$.

\subsection{Matching to the leading order of $\chi$PT}
The dim-6 quark-neutrino operators can be matched onto the meson-lepton interactions through the $\chi$PT formalism by treating the lepton currents together with the accompanied Wilson coefficients as proper external sources. The QCD-like Lagrangian with external sources for the first three light quarks ($q=u,d,s$) can be described as
\begin{eqnarray}
\mathcal{L}_{\rm QCD}=\mathcal{L}_{\text{QCD}}^{m_q=0}+\overline{q_L}l_\mu \gamma^\mu q_L+\overline{q_R}r_\mu \gamma^\mu q_R+\left[\overline{q_L}(s-ip)q_R+\overline{q_L}(t_l^{\mu\nu}\sigma_{\mu\nu})q_R+h.c.\right],
\end{eqnarray}
where the flavor space $3\times$3 matrices $\{l_\mu=l_\mu^\dagger,~r_\mu=r_\mu^\dagger,~s=s^\dagger,~p=p^\dagger,~t_r^{\mu\nu}=t_l^{\mu\nu\dagger}\}$ are the external sources related with the corresponding quark currents. One can extract the relevant external sources from the above dim-6 effective operators. On the other hand, based on Weinberg's power-counting scheme, the most general chiral Lagrangian can be expanded according to the momentum $p$ and quark mass.
The chiral Lagrangian with external sources at leading order reads~\cite{Gasser:1983yg,Gasser:1984gg}
\begin{eqnarray}
\mathcal{L}_{p^2}=\frac{F_0^2}{4}{\Tr}\left(D_\mu U (D^\mu U)^\dagger \right)+\frac{F_0^2}{4}{\Tr} \left(\chi U^\dagger +U\chi^\dagger \right),
\label{eq:L2}
\end{eqnarray}
where $U$ is the standard matrix for the Nambu-Goldstone bosons
\begin{eqnarray}
U={\rm exp}\Big({i\Phi\over F_0}\Big), \ \ \Phi=\left(
  \begin{array}{ccc}
    \pi^0+{\eta\over \sqrt{3}} &\sqrt{2} \pi^+ & \sqrt{2} K^+ \\
    \sqrt{2}\pi^- & -\pi^0+{\eta\over \sqrt{3}}& \sqrt{2}K^0 \\
    \sqrt{2}K^- & \sqrt{2}\bar{K}^0 & -{2\over \sqrt{3}}\eta \\
  \end{array}
\right) \; ,
\end{eqnarray}
with the constant $F_0$ being referred to the pion decay constant in the chiral limit.
The covariant derivative of $U$ and $\chi$ are expressed in terms of the external sources
\begin{align}
	D_\mu U&=\partial_\mu U-i l_\mu U+i U r_\mu,&
	\chi& =2B(s-ip),&
	\chi^\dagger&=2B(s+ip),
\end{align}
where the constant $B$ is related to the quark condensate and $F_0$ by $B=-\langle\bar{q}q\rangle_0/(3F_0^2)$. For the later numerical estimation, we take
$F_0=87~\text{MeV}$~\cite{Colangelo:2003hf} and $B\approx 2.8~\text{GeV}$~\cite{Cirigliano:2017djv,Patrignani:2016xqp}.
The Nambu-Goldstone bosons parameterized by $U$ and the (pseudo-)scalar sources $\chi$ transform as $U\to L U R^\dagger$ and $\chi\to L \chi R^\dagger$, where $L$ ($R$) is $SU(3)_L$ ($SU(3)_R$) transformation.

By inspecting the dim-6 operators related to the $s\rightarrow d\nu\widehat{\nu}$ transition (the symbol ~`~$\widehat{}$~'~ here indicating the neutrino pair can be either LNC $\nu\bar{\nu}$ or LNV $\nu\nu$\footnote{Below we use $K\to \pi \nu\bar{\nu}$ to generally denote the experimental processes. $K\to \pi \nu\widehat{\nu}$ appears when both $\nu\bar{\nu}$ and $\nu\nu$ final states can occur in the analytical expressions of the theoretical calculation unless a LNC or LNV process is specified in our discussion.}), we find that only the LNC operators $\calO_{d\nu 1}^V,~\calO_{d\nu 2}^V$ and LNV operators $\calO_{d\nu 1}^S,~\calO_{d\nu 2}^S$ can contribute to the leading order chiral Lagrangian. The tensor operator $\calO_{d\nu}^T$ only contributes to the next-to-leading order chiral Lagrangian at $\mathcal{O}(p^4)$. They lead to the following external sources
\begin{eqnarray}
(l^\mu)_{sd}&=&C_{d\nu1}^{V,sd\alpha\beta}(\overline{\nu_\alpha} \gamma^\mu\nu_\beta),
\\
(l^\mu)_{ds}&=&C_{d\nu1}^{V,ds\alpha\beta}(\overline{\nu_\alpha} \gamma^\mu\nu_\beta),
\\
(r^\mu)_{sd}&=&C_{d\nu2}^{V,sd\alpha\beta}(\overline{\nu_\alpha} \gamma^\mu\nu_\beta),
\\
(r^\mu)_{ds}&=&C_{d\nu2}^{V,ds\alpha\beta}(\overline{\nu_\alpha} \gamma^\mu\nu_\beta),
\\
(s+ip)_{sd}&=&C_{d\nu1}^{S,sd\alpha\beta}(\overline{\nu^C_\alpha} \nu_\beta)+C_{d\nu2}^{S,ds\alpha\beta*}(\overline{\nu_\alpha} \nu_\beta^C),
\\
(s+ip)_{ds}&=&C_{d\nu1}^{S,ds\alpha\beta}(\overline{\nu^C_\alpha} \nu_\beta)+C_{d\nu2}^{S,sd\alpha\beta*}(\overline{\nu_\alpha} \nu_\beta^C),
\\
(s-ip)_{sd}&=&C_{d\nu1}^{S,ds\alpha\beta*}(\overline{\nu_\alpha} \nu_\beta^C)+C_{d\nu2}^{S,sd\alpha\beta}(\overline{\nu^C_\alpha} \nu_\beta),
\\
(s-ip)_{ds}&=&C_{d\nu1}^{S,sd\alpha\beta*}(\overline{\nu_\alpha} \nu_\beta^C)+C_{d\nu2}^{S,ds\alpha\beta}(\overline{\nu^C_\alpha} \nu_\beta) .
\end{eqnarray}

After expanding $U$, i.e. $U=1+i{\Phi\over F_0}+{1\over 2F_0^2}(i\Phi)^2+\cdots$ and the insertion of the above external sources into the Lagrangian in Eq.~(\ref{eq:L2}), we obtain the effective Lagrangians for $K^0\to \pi^0\nu\widehat{\nu}$ and $K^+\to \pi^+\nu\widehat{\nu}$ at the leading order
\begin{eqnarray}\nonumber
\mathcal{L}_{K^0\to \pi^0\nu\widehat{\nu}}&=&
{B\over 2\sqrt{2}}\Big[
\left(C_{d\nu1}^{S,sd\alpha\beta}+C_{d\nu2}^{S,sd\alpha\beta}\right)(\overline{\nu^C_\alpha}\nu_\beta)
+\left(C_{d\nu1}^{S,ds\alpha\beta*}+C_{d\nu2}^{S,ds\alpha\beta*}\right)(\overline{\nu_\alpha}\nu_\beta^C)\Big]K^0\pi^0
\\
&&
-{i\over 2\sqrt{2}}\left(C_{d\nu 1}^{V,sd\alpha\beta}+C_{d\nu 2}^{V,sd\alpha\beta}\right)(\overline{\nu_\alpha}\gamma^\mu\nu_\beta)(K^0 \partial_\mu \pi^0-\partial_\mu K^0 \pi^0)\; ,
\\\nonumber
\mathcal{L}_{K^+\to \pi^+\nu\widehat{\nu}}&=&
-{B\over 2}\Big[
\left(C_{d\nu1}^{S,sd\alpha\beta}+C_{d\nu2}^{S,sd\alpha\beta}\right)(\overline{\nu^C_\alpha}\nu_\beta)
+\left(C_{d\nu1}^{S,ds\alpha\beta*}+C_{d\nu2}^{S,ds\alpha\beta*}\right)(\overline{\nu_\alpha}\nu_\beta^C)\Big]K^+\pi^-
\\
&&
+{i\over 2}\left(C_{d\nu 1}^{V,sd\alpha\beta}+C_{d\nu 2}^{V,sd\alpha\beta}\right)(\overline{\nu_\alpha}\gamma^\mu\nu_\beta)(K^+ \partial_\mu \pi^--\partial_\mu K^+ \pi^-)\; .
\end{eqnarray}
The above Lagrangians fit to the relation
\begin{eqnarray}
{\langle\pi^0| \mathcal{L}_{K^0\to \pi^0\nu\widehat{\nu}}|K^0\rangle \over \langle\pi^-|\mathcal{L}_{K^+\to \pi^+\nu\widehat{\nu}}|K^+\rangle} =-\frac{1}{\sqrt{2}} \; .
\label{eq:isospin1over2}
\end{eqnarray}
This relation is the result of the transition operators that change isospin by $1/2$.
By neglecting the small CP violation in $K^0-\bar K^0$ mixing, for the $K_L\rightarrow\pi^0$ transition, the relevant effective Lagrangian becomes
\begin{eqnarray}
\mathcal{L}_{K_L\to \pi^0\nu\widehat{\nu}}&=&
{B\over 4}\Big[
\left(C_{d\nu1}^{S,sd\alpha\beta}+C_{d\nu2}^{S,sd\alpha\beta}
+C_{d\nu1}^{S,ds\alpha\beta}+C_{d\nu2}^{S,ds\alpha\beta}\right)(\overline{\nu^C_\alpha}\nu_\beta) \\
&&+\left(C_{d\nu1}^{S,ds\alpha\beta*}+C_{d\nu2}^{S,ds\alpha\beta*}
+C_{d\nu1}^{S,sd\alpha\beta*}+C_{d\nu2}^{S,sd\alpha\beta*}\right)(\overline{\nu_\alpha}\nu_\beta^C)\Big]K_L\pi^0 \nonumber \\
&&
-{i\over 4}\left(C_{d\nu 1}^{V,sd\alpha\beta}+C_{d\nu 2}^{V,sd\alpha\beta}-C_{d\nu1}^{V,ds\alpha\beta}-C_{d\nu 2}^{V,ds\alpha\beta}\right)(\overline{\nu_\alpha}\gamma^\mu\nu_\beta)(K_L \partial_\mu \pi^0-\partial_\mu K_L \pi^0) \; , \nonumber
\end{eqnarray}
where the flavor indices $\alpha,\beta$ are summed over all three neutrino generations.
Note that the Wilson coefficients for the scalar operators are symmetric in the neutrino flavor indices. From the effective Lagrangian we derive the branching ratios for the decays $K_L\to \pi^0\nu\widehat{\nu}$ and $K^+\to \pi^+\nu\widehat{\nu}$
\begin{eqnarray}\nonumber
	\mathcal{B}_{K_L\rightarrow\pi^0\nu\widehat{\nu}}&=&J_1^{K_L} \sum_{\alpha\leq \beta} \left(1-{1\over2}\delta_{\alpha\beta}\right)  \left|C_{d\nu1}^{S,sd\alpha\beta}+C_{d\nu2}^{S,sd\alpha\beta}
+C_{d\nu1}^{S,ds\alpha\beta}+C_{d\nu2}^{S,ds\alpha\beta}\right|^2
\\\label{brkl}
&&+J_2^{K_L}\sum_{\alpha, \beta} \left|C_{d\nu 1}^{V,sd\alpha\beta}+C_{d\nu 2}^{V,sd\alpha\beta}-C_{d\nu1}^{V,ds\alpha\beta}-C_{d\nu 2}^{V,ds\alpha\beta}\right|^2,
\\\nonumber
\mathcal{B}_{K^+\rightarrow\pi^+\nu\widehat\nu}&=&J_1^{K^+}\sum_{\alpha\leq \beta} \left(1-{1\over2}\delta_{\alpha\beta}\right)
  \left(\left|C_{d\nu1}^{S,sd\alpha\beta}+C_{d\nu2}^{S,sd\alpha\beta}\right|^2+\left|C_{d\nu1}^{S,ds\alpha\beta}+C_{d\nu2}^{S,ds\alpha\beta}\right|^2\right)
\\\label{brkp}
&&+J_2^{K^+}\sum_{\alpha, \beta}  \left|C_{d\nu 1}^{V,sd\alpha\beta}+C_{d\nu 2}^{V,sd\alpha\beta}\right|^2.
\end{eqnarray}
The details of the calculation are collected in Appendix~\ref{sec:Kpinunu}.
The $J$ functions parameterize the kinematics of the three-body decay and are defined as
\begin{eqnarray}
	J_1^{K_L}&=&{1 \over \Gamma_{K_L}^{\text{Exp}}}{B^2\over 2^9\pi^3m_{K_L}^3}\int ds\, s\left((m_{K_L}^2+m_{\pi^0}^2-s)^2-4m_{K_L}^2m_{\pi^0}^2\right)^{1/2}=40.4 G_F^{-2},
\\
J_2^{K_L}&=&{1 \over \Gamma_{K_L}^{\text{Exp}}}{1\over 3\cdot 2^{11}\pi^3m_{K_L}^3}\int ds \left((m_{K_L}^2+m_{\pi^0}^2-s)^2-4m_{K_L}^2m_{\pi^0}^2\right)^{3/2}=0.247 G_F^{-2},
\\
J_1^{K^+}&=&{1 \over \Gamma_{K^+}^{\text{Exp}}}{ B^2\over 2^8\pi^3m_{K^+}^3}\int ds\, s\left((m_{K^+}^2+m_{\pi^+}^2-s)^2-4m_{K^+}^2m_{\pi^+}^2\right)^{1/2}=17.9 G_F^{-2},
\\
J_2^{K^+}&=&{1 \over \Gamma_{K^+}^{\text{Exp}}}{1\over 3\cdot 2^9\pi^3m_{K^+}^3}\int ds \left((m_{K^+}^2+m_{\pi^+}^2-s)^2-4m_{K^+}^2m_{\pi^+}^2\right)^{3/2}=0.22 G_F^{-2}\; ,
\end{eqnarray}
where $m_{K_L}(m_{K^+})$ and $\Gamma_{K_L}^{\text{Exp}}(\Gamma_{K^+}^{\text{Exp}})$ denote the physical mass and decay width of $K_L(K^+)$, respectively. $m_{\pi^0}(m_{\pi^+})$ is the mass of $\pi^0(\pi^+)$, $G_F$ is the Fermi constant and $s$ is the invariant squared mass of the final-state neutrino pair.
From the hermiticity of the effective Lagrangian and the Cauchy-Schwarz inequality we derive the following relations for the Wilson coefficients in LEFT
\begin{eqnarray}
\left|C_{d\nu1}^{S,sd\alpha\beta}+C_{d\nu2}^{S,sd\alpha\beta}
+C_{d\nu1}^{S,ds\alpha\beta}+C_{d\nu2}^{S,ds\alpha\beta}\right|^2
&\leq&
\left| |C_{d\nu1}^{S,sd\alpha\beta}+C_{d\nu2}^{S,sd\alpha\beta}|
+|C_{d\nu1}^{S,ds\alpha\beta}+C_{d\nu2}^{S,ds\alpha\beta}|\right|^2 \nonumber \\
&\leq&
2\left( \left|C_{d\nu1}^{S,sd\alpha\beta}+C_{d\nu2}^{S,sd\alpha\beta}\right|^2+\left|C_{d\nu1}^{S,ds\alpha\beta}+C_{d\nu2}^{S,ds\alpha\beta}\right|^2 \right) \; , \nonumber\\
\\
\left|C_{d\nu 1}^{V,sd\alpha\beta}+C_{d\nu 2}^{V,sd\alpha\beta}-C_{d\nu1}^{V,ds\alpha\beta}-C_{d\nu 2}^{V,ds\alpha\beta}\right|^2&\leq&
\left| |C_{d\nu 1}^{V,sd\alpha\beta}+C_{d\nu 2}^{V,sd\alpha\beta}|+|C_{d\nu1}^{V,ds\alpha\beta}+C_{d\nu 2}^{V,ds\alpha\beta}|\right|^2 \nonumber
\\
&\leq& 2\left( \left|C_{d\nu1}^{V,sd\alpha\beta}+C_{d\nu2}^{V,sd\alpha\beta}\right|^2
+\left|C_{d\nu1}^{V,sd\beta\alpha}+C_{d\nu2}^{V,sd\beta\alpha}\right|^2 \right) \; . \nonumber \\
\end{eqnarray}
Note that in the second inequality we used $C_{d\nu1/2}^{V,ds\alpha\beta*}=C_{d\nu1/2}^{V,sd\beta\alpha}$ from the fact that the vector operator itself is hermitian. If we sum over neutrino flavors, the second relation above turns out to be
\begin{eqnarray}
\sum_{\alpha,\beta}\left|C_{d\nu 1}^{V,sd\alpha\beta}+C_{d\nu 2}^{V,sd\alpha\beta}-C_{d\nu1}^{V,ds\alpha\beta}-C_{d\nu 2}^{V,ds\alpha\beta}\right|^2\leq 4 \sum_{\alpha,\beta}\left|C_{d\nu1}^{V,sd\alpha\beta}+C_{d\nu2}^{V,sd\alpha\beta}\right|^2 \; .
\end{eqnarray}

Based on the above inequalities, the branching ratios in Eq.~(\ref{brkl}) and Eq.~(\ref{brkp}) lead to
\begin{eqnarray}
{\mathcal{B}_{K_L\rightarrow\pi^0\nu\widehat{\nu}} \over \mathcal{B}_{K^+\rightarrow\pi^+\nu\widehat{\nu}}} \leq
{4J_2^{K_L}+2J_1^{K_L}\epsilon \over J_2^{K^+}+J_1^{K^+}\epsilon}
\lesssim 4.5\; ,
\label{GNbound}
\end{eqnarray}
where $\epsilon$ is defined by
\begin{eqnarray}
\epsilon={\sum_{\alpha\leq \beta} \left(1-{1\over2}\delta_{\alpha\beta}\right)
\left(\left|C_{d\nu1}^{S,sd\alpha\beta}+C_{d\nu2}^{S,sd\alpha\beta}\right|^2+\left|C_{d\nu1}^{S,ds\alpha\beta}+C_{d\nu2}^{S,ds\alpha\beta}\right|^2\right)\over \sum_{\alpha,\beta}\left|C_{d\nu1}^{V,sd\alpha\beta}+C_{d\nu2}^{V,sd\alpha\beta}\right|^2}\geq0 \; .
\end{eqnarray}
The upper bound on the ratio of branching ratios in Eq.~\eqref{GNbound} holds independent of the value of $\epsilon\geq 0$, because $4J_2^{K_L}/J_2^{K^+}\approx 4.49$ and $2J_1^{K_L}/J_1^{K^+}\approx 4.51$ agree to two significant figures.
This result is nothing but the Grossman-Nir bound~\cite{Grossman:1997sk} expected from the isospin relation in Eq.~(\ref{eq:isospin1over2}) and the CP-conserving limit for neutral Kaon system. The numerical value slightly differs from the standard G-N bound value of 4.3, because we do not consider isospin breaking and electroweak correction effects beyond the mass difference in the phase space integration~\cite{Marciano:1996wy}.
We obtain the G-N bound from the matching of LEFT to $\chi$PT. Thus, as expected, the Grossman-Nir bound holds for dim-6 LEFT operators in leading order $\chi$PT.

\subsection{Dim-6 tensor operators and dim-7 operators in the chiral Lagrangian}
For the tensor currents in Eqs.~(\ref{tensor}), we have to go beyond the leading order of chiral Lagrangian. The relevant $\mathcal{O}(p^4)$ Lagrangian at next-to-leading order is~\cite{Cata:2007ns}
\begin{eqnarray}\label{eq:Lp4}
\mathcal{L}_{p^4}^T \supset i\Lambda_2 {\Tr}\left(t_l^{\mu\nu}(D_\mu U)^\dagger U (D_\nu U)^\dagger+t_r^{\mu\nu}D_\mu UU^\dagger D_\nu U\right),
\end{eqnarray}
where $\Lambda_2$ denotes the low-energy constant. In terms of the dim-6 tensor operator $\calO_{d\nu}^T$, the relevant tensor sources are
\begin{align}
(t_r^{\mu\nu})_{ds}&=C_{d\nu}^{T,ds\alpha\beta}(\overline{\nu_\alpha^C}\sigma_{\mu\nu}\nu_\beta),
		   &
(t_r^{\mu\nu})_{sd}&=C_{d\nu}^{T,sd\alpha\beta}(\overline{\nu_\alpha^C}\sigma_{\mu\nu}\nu_\beta),
\\
(t_l^{\mu\nu})_{ds}&=C_{d\nu}^{T,sd\alpha\beta*}(\overline{\nu_\alpha}\sigma_{\mu\nu}\nu_\beta^C),
		   &
(t_l^{\mu\nu})_{sd}&=C_{d\nu}^{T,ds\alpha\beta*}(\overline{\nu_\alpha}\sigma_{\mu\nu}\nu_\beta^C).
\end{align}
After inserting these external sources into the $\mathcal{O}(p^4)$ Lagrangian \eqref{eq:Lp4}, we obtain the following interactions for $K\to \pi\nu\nu$ transitions
\begin{eqnarray}\nonumber
\mathcal{L}_{p^4}^T
&\supset& i {\Lambda_2 \over \sqrt{2} F_0^2}C_{d\nu}^{T,sd\alpha\beta}\left(\sqrt{2} \partial_{[\mu}K^+\partial_{\nu]}\pi^-- \partial_{[\mu}K^0\partial_{\nu]}\pi^0\right)(\overline{\nu_\alpha^C}\sigma_{\mu\nu}\nu_\beta)
\\
&&
+i {\Lambda_2 \over \sqrt{2} F_0^2}C_{d\nu}^{T,ds\alpha\beta*}\left(\sqrt{2} \partial_{[\mu}K^+\partial_{\nu]}\pi^-- \partial_{[\mu}K^0\partial_{\nu]}\pi^0\right)(\overline{\nu_\alpha}\sigma_{\mu\nu}\nu_\beta^C)+h.c.,
\end{eqnarray}
where $\partial_{[\mu}A\partial_{\nu]}B=\partial_\mu A\partial_\nu B-\partial_\nu A\partial_\mu B$.

We also investigate dim-7 tensor operators in LEFT. There happens to be only one such operator related with the transition $K\to \pi\nu\bar{\nu}$ under consideration, that is
\begin{eqnarray}
\calO_{d\nu, \text{dim-7}}^{T}&=&(\overline{d_L}\sigma_{\mu\nu}d_R)(\overline{\nu}\gamma^{[\mu}i\overleftrightarrow{\partial}^{\nu]}\nu)+h.c. \; ,
\end{eqnarray}
which leads to the tensor sources to be
\begin{align}
(t_l^{\mu\nu})_{ds}&=C^{T, ds\alpha\beta}_{d\nu, \text{dim-7}}(\overline{\nu_\alpha}\gamma^{[\mu}i\overleftrightarrow{\partial}^{\nu]}\nu_\beta),
		   &
(t_l^{\mu\nu})_{sd}&=C^{T, sd\alpha\beta}_{d\nu, \text{dim-7}}(\overline{\nu_\alpha}\gamma^{[\mu}i\overleftrightarrow{\partial}^{\nu]}\nu_\beta),
\\
(t_r^{\mu\nu})_{ds}&=C^{T, sd\beta\alpha*}_{d\nu, \text{dim-7}}(\overline{\nu_\alpha}\gamma^{[\mu}i\overleftrightarrow{\partial}^{\nu]}\nu_\beta),
			  &
(t_r^{\mu\nu})_{sd}&=C^{T, ds\beta\alpha*}_{d\nu, \text{dim-7}}(\overline{\nu_\alpha}\gamma^{[\mu}i\overleftrightarrow{\partial}^{\nu]}\nu_\beta).
\end{align}
By analogy we expand the $\mathcal{O}(p^4)$ Lagrangian \eqref{eq:Lp4} to obtain the interactions with mesons
\begin{eqnarray}\label{dim7tensor}
\mathcal{L}_{p^4}^T
\supset {i\Lambda_2\over \sqrt{2}F_0^2} (C^{T, sd\alpha\beta}_{d\nu, \text{dim-7}}+C^{T, ds\beta\alpha*}_{d\nu, \text{dim-7}})\left( \sqrt{2}\partial_{[\mu }K^+ \partial_{\nu]}\pi^- -\partial_{[\mu }K^0 \partial_{\nu]}\pi^0 \right)(\overline{\nu_\alpha}\gamma^{[\mu}i\overleftrightarrow{\partial}^{\nu]}\nu_\beta)+h.c. \; .\nonumber\\
\end{eqnarray}
One can see that, for the next-to-leading order chiral Lagrangian with dim-6 and dim-7 tensor operators in LEFT, the isospin relation in Eq.~(\ref{eq:isospin1over2}) and thus the Grossman-Nir bound still hold.
Note that the Eq.~\eqref{dim7tensor} vanishes for massless neutrinos. This implies that the non-zero contribution from dim-7 tensor operator appears at $\calO(p^6)$ level, and therefore is further suppressed by additional $p^2/\Lambda_\chi^2$ factor.
In addition, there are also two dim-7 vector-like LNV operators related to $K\to\pi\nu\nu$, which we list for completeness
\begin{align}
\calO_{d\nu1, \text{dim-7}}^{V}&=(\overline{d_L}\gamma_\mu d_L)(\overline{\nu^C}i\overleftrightarrow{\partial}^{\mu}\nu)+h.c. \; ,
			       &
\calO_{d\nu2, \text{dim-7}}^{V}&=(\overline{d_R}\gamma_\mu d_R)(\overline{\nu^C}i\overleftrightarrow{\partial}^{\mu}\nu)+h.c. \; .
\end{align}
These dim-7 operators are suppressed by $p/m_W$ compared with dim-6 operators and, like the above tensor operators, lead to sub-leading contributions. Thus, we neglect them in the following calculation and restrict us to only consider the scalar and vector dim-6 operators in LEFT.

\section{Matching to the SMEFT}
\label{sec:SMEFT}

Next we need to match the Wilson coefficients relevant for the $K\to \pi\nu\widehat{\nu}$ processes in LEFT to those in SMEFT at the electroweak scale $\Lambda_{\text{EW}}$, by integrating out heavy SM particles.
First of all, the SM contribution to the $s\to d$ transition occurs at loop-level and it only matches to the LEFT operator $\calO_{d\nu 1}^V$ by integrating out heavy SM particles at the one-loop level~\cite{Buchalla:1993wq,Buchalla:1998ba}
\begin{eqnarray}\label{eq:SMcontribution}
C_{d\nu 1,\text{SM}}^{V,sd\alpha\beta}&=&
{G_{F}\over \sqrt{2}  }{2\alpha_{\text{EM}} \over \pi s_W^2} \delta_{\alpha\beta}\left(V_{cs}^*V_{cd} X^{\alpha}+V_{ts}^*V_{td} X_{t}\right),
\\
C_{d\nu 1,\text{SM}}^{V,ds\alpha\beta}&=&C_{d\nu 1,\text{SM}}^{V,sd\beta\alpha*}, \\
C_{d\nu 2,\text{SM}}^{V,sd\alpha\beta}&=&0 \; ,
\end{eqnarray}
where the loop function $X^\alpha/X_t$ can be found in Refs.~\cite{Buchalla:1993wq,Buchalla:1998ba} and higher order corrections are given in Ref.~\cite{Brod:2010hi}. We take the central values for CKM elements from CKMfitter~\cite{CKMfitter}, $X_t$ and $X^\alpha$ from Ref.~\cite{He:2018uey}, and the rest from the PDG book~\cite{Tanabashi:2018oca}. Then, to the leading order in $\chi$PT,
the analytical expressions in Eqs.~\eqref{brkl} and \eqref{brkp} with the Wilson coefficients in Eqs.~\eqref{eq:SMcontribution} predict the branching ratios of Kaon semi-invisible decays in the SM
\begin{align}
\mathcal{B}_{K_L\rightarrow\pi^0\nu\bar\nu}^{\text{SM}}&=2.99\times 10^{-11}, &
\mathcal{B}_{K^+\rightarrow\pi^+\nu\bar\nu}^{\text{SM}}&=8.31\times 10^{-11},
\end{align}
which are consistent with SM predictions quoted in the literature~\cite{Buras:2006gb,Brod:2010hi,Buras:2015qea}.

Secondly, the dim-6 SMEFT operators in the Warsaw basis~\cite{Grzadkowski:2010es} and the dim-7 SMEFT operators in the basis given in Refs.~\cite{Liao:2016hru,Lehman:2014jma} can induce the operators in the LEFT by integrating out the SM particles at tree-level.
The LNC operators $\calO^V_{d\nu1/2}$ are obtained through matching with the dim-6 SMEFT operators in addition to the SM contribution in Eq.~(\ref{eq:SMcontribution}).
To linear order in the SMEFT Wilson coefficients, the matching results are
\begin{eqnarray}\nonumber
C_{d\nu 1,\text{dim-6}}^{V,sd\alpha\beta}&=&D_{xs}^*D_{yd}\left(
C_{l q}^{(1),\alpha\beta xy}-C_{l q}^{(3),\alpha\beta xy}
+\left(C_{Hq}^{(1),xy}+C_{Hq}^{(3),xy}\right)\delta_{\alpha\beta}
\right)
\\\label{cdnu1}
&\approx&C_{l q}^{(1),\alpha\beta 21}-C_{l q}^{(3),\alpha\beta 21}
+\left( C_{Hq}^{(1),21}+C_{Hq}^{(3),21}\right)\delta_{\alpha\beta},
\\\nonumber
C_{d\nu 1,\text{dim-6}}^{V,ds\alpha\beta}&=&C_{d\nu 1}^{V,sd\beta\alpha*}=D_{xd}^*D_{ys}\left(
C_{l q}^{(1), \alpha\beta xy}-C_{l q}^{(3),\alpha\beta xy}
+\left(C_{Hq}^{(1),xy}+C_{Hq}^{(3),xy}\right)\delta_{\alpha\beta}
\right)
\\
&\approx&C_{l q}^{(1), \alpha\beta 12}-C_{l q}^{(3),\alpha\beta 12}
+\left( C_{Hq}^{(1),12}+C_{Hq}^{(3),12}\right)\delta_{\alpha\beta},
\\
C_{d\nu 2,\text{dim-6}}^{V,sd\alpha\beta}&=&
C_{l d}^{\alpha\beta 21}+C_{Hd}^{21}\delta_{\alpha\beta},
\\\label{cdnu2}
C_{d\nu 2,\text{dim-6}}^{V,ds\alpha\beta}&=&C_{d\nu 2}^{V,sd\beta\alpha*}=
C_{l d}^{\alpha\beta 12}+C_{Hd}^{12}\delta_{\alpha\beta},
\end{eqnarray}
where $D$ is the unitary matrix transforming left-handed down-type quarks between flavor
$d_L^\prime$ and mass eigenstates $d_L$, $d_L=D d_L^\prime$.
We choose $D$ to be approximately the identity matrix and neglect its
effect in the following, i.e.~the weak interaction eigenstates are
the same as the mass eigenstates and the mixing originates from the
up-type quarks. The convention for the Wilson coefficients is taken
from Ref.~\cite{Grzadkowski:2010es}, with the corresponding SMEFT
operators being
\begin{align}\label{dim61}
\calO_{lq}^{(1)}&=({\overline{L}\gamma^\mu L})(\overline{Q}\gamma_\mu Q),
		&
\calO_{lq}^{(3)}&=(\overline{L}\gamma^\mu \sigma^I L)(\overline{Q}\gamma_\mu  \sigma^I Q),
		&
\calO_{ld}&=(\overline{L}\gamma^\mu L)(\overline{d}\gamma_\mu d),
\\
\calO_{Hq}^{(1)}&=(H^\dagger i \overleftrightarrow{D_\mu}H)(\overline{Q}\gamma_\mu Q),
		&
\calO_{Hq}^{(3)}&=(H^\dagger i \overleftrightarrow{D_\mu^I}H)(\overline{Q}\gamma_\mu \sigma^I Q),
		&\label{dim66}
\calO_{Hd}&=(H^\dagger i \overleftrightarrow{D_\mu}H)(\overline{d}\gamma_\mu d)
\;.
\end{align}
The $\sigma^I$ are the Pauli matrices, and $H^\dagger i\overleftrightarrow{D_\mu^I}H=iH^\dagger \sigma^I D_\mu H- i(D_\mu H)^\dagger \sigma^I H$.

For the LNV operators $\calO^S_{d\nu1/2}$ the leading contribution comes from dim-7 SMEFT operators, since dim-6 operators in SMEFT do not violate lepton number by two units and the dim-5 Weinberg operator is strongly constrained from neutrino masses and only indirectly contributes to the LEFT operators.
The only dim-7 SMEFT operator which induces $\calO^S_{d\nu1}$ at tree-level is~\cite{Liao:2016hru}
\begin{eqnarray}\label{dim7operator}
\mathcal{O}_{\overline{d}LQLH1}=\epsilon_{ij}\epsilon_{mn}(\overline{d}L^i)(\overline{Q^{C,j}}L^m)H^n,
\end{eqnarray}
with the matching result for the Wilson coefficients at the electroweak scale
\begin{eqnarray}\label{csds1}
C_{d\nu 1, \text{dim-7}}^{S,ds\alpha\beta}&=&
-{\sqrt{2}\over 8}vD_{xs}\left(C_{\bar{d}LQLH1}^{1\alpha x\beta}+C_{\bar{d}LQLH1}^{1\beta x\alpha} \right)
\approx
-{\sqrt{2}\over 8}v\left(C_{\bar{d}LQLH1}^{1\alpha 2\beta}+C_{\bar{d}LQLH1}^{1\beta 2\alpha} \right),
\\\label{cssd1}
C_{d\nu 1, \text{dim-7}}^{S,sd\alpha\beta}&=&
-{\sqrt{2}\over 8}vD_{xd}\left(C_{\bar{d}LQLH1}^{2\alpha x\beta}+C_{\bar{d}LQLH1}^{2\beta x\alpha} \right)\approx
-{\sqrt{2}\over 8}v\left(C_{\bar{d}LQLH1}^{2\alpha 1\beta}+C_{\bar{d}LQLH1}^{2\beta 1\alpha} \right).
\end{eqnarray}
We use subscripts $1$, $2$, and $x$ to represent the SM quark generation.
The indices $\alpha$ or $\beta$ denote the SM lepton flavor. As the operator violates quark flavor, the contribution to neutrino masses is suppressed and does not pose a stringent constraint.
Note that the $\calO^S_{d\nu 2}$ operator cannot be induced at tree-level from SMEFT.
In the following we derive constraints on the SMEFT operators from $K\to\pi\nu\bar\nu$ and compare to the existing measurements of other related processes.

A brief comment on renormalization group corrections in LEFT is in order. As neutrinos neither couple to gluons nor photons, we only have to consider QCD corrections. Due to the QCD Ward identity, there are no QCD corrections to the vector operators at one-loop order and the running of the scalar operator can be simply obtained by noting that $m_f \bar f P_{L,R} f$ is invariant under QCD renormalization group corrections. Hence, the running of the scalar operator can be directly related to the QCD correction to the quark masses, $C_S(\mu) = C_S(m_W) m_q(m_W) /m_q(\mu)$.

\section{Implication of $K\to \pi \nu \bar{\nu}$ for new physics and other constraints}
\label{sec:Implication}

In this section, based on the above LEFT coefficients in the leading-order chiral Lagrangian and the matching to SMEFT, we evaluate the constraints on new physics above the electroweak scale from the $K\to \pi \nu \bar{\nu}$ measurements and other rare decays. According to the decay branching ratios in Eqs.~(\ref{brkl}) and (\ref{brkp}), to the leading order of the chiral Lagrangian, both vector and scalar LEFT operators contribute to the decays $K\to \pi \nu \widehat{\nu}$. They correspond to dim-6 LNC operators and one dim-7 LNV operator in the SMEFT, respectively. We will separately discuss the constraints on them below.

\subsection{Constraint on the LNC operators}

From the branching ratios in Eqs.~(\ref{brkl}-\ref{brkp}), and the matching results in Eqs.~(\ref{cdnu1}-\ref{cdnu2}), we split the contributions to the amplitude into the SM part given in Eq.~(\ref{eq:SMcontribution}) and the NP part as follows
\begin{eqnarray}\label{dim6contribution}
C_{d\nu 1}^{V,sd\alpha\beta}+C_{d\nu 2}^{V,sd\alpha\beta}=
C_{d\nu 1,\text{SM}}^{V,sd \alpha\beta}+C_{d\nu,\text{dim-6}}^{V,sd\alpha\beta} \; ,
\label{SManddim6}
\end{eqnarray}
where the NP part is the linear combination of the Wilson coefficients of dim-6 LNC operators in the SMEFT in Eqs.~(\ref{cdnu1}-\ref{cdnu2})
\begin{eqnarray}
C_{d\nu,\text{dim-6}}^{V,sd\alpha\beta}&=&C_{d\nu 1,\text{dim-6}}^{V,sd\alpha\beta}+C_{d\nu 2,\text{dim-6}}^{V,sd\alpha\beta} \nonumber \\
&\approx& C_{l q}^{(1),\alpha\beta 21}+C_{l d}^{\alpha\beta 21}
-C_{l q}^{(3),\alpha\beta 21}
+\left( C_{Hq}^{(1),21}+C_{Hq}^{(3),21}+C_{Hd}^{21}\right)\delta_{\alpha\beta}\; .
\end{eqnarray}
Taking the splitting in Eq.~(\ref{dim6contribution}) and the experimental results in Eqs.~(\ref{koto},~\ref{na62}), we find
\begin{eqnarray}
\mathcal{B}_{K_L\rightarrow\pi^0\nu\bar\nu}&=&J_2^{K_L}\sum_{\alpha, \beta}\left|
2{\rm Im}[C_{d\nu 1,\text{SM}}^{V,sd \alpha\beta}]+C_{d\nu,\text{dim-6}}^{V,sd\alpha\beta}-C_{d\nu,\text{dim-6}}^{V,ds\alpha\beta}\right|^2\in [0.4\times10^{-9},~6.2\times 10^{-9}],
\\
\mathcal{B}_{K^+\rightarrow\pi^+\nu\bar\nu}
&=&J_2^{K^+} \sum_{\alpha, \beta} \left|C_{d\nu 1,\text{SM}}^{V,sd \alpha\beta}+C_{d\nu,\text{dim-6}}^{V,sd\alpha\beta}\right|^2 < 2.44\times10^{-10} \; .
\end{eqnarray}
The following generic relations can be immediately obtained
\begin{eqnarray}\label{kotobound}
&&1.62\times 10^{-9} G_F^{2} < \sum_{\alpha, \beta}\left|2{\rm Im}[C_{d\nu 1,\text{SM}}^{V,sd \alpha\beta}]+C_{d\nu,\text{dim-6}}^{V,sd\alpha\beta}-C_{d\nu,\text{dim-6}}^{V,ds\alpha\beta}\right|^2 < 2.51\times 10^{-8} G_F^{2} \; ,
\\\label{na62bound}
&&\sum_{\alpha, \beta}\left|C_{d\nu 1,\text{SM}}^{V,sd \alpha\beta}+C_{d\nu,\text{dim-6}}^{V,sd\alpha\beta}\right|^2 < 1.11\times 10^{-9} G_F^{2} \; .
\end{eqnarray}

Now we first consider the lepton-flavor-conserving (LFC) case and ignore the lepton-flavor-violating (LFV) contributions for the time being. After electroweak symmetry breaking, the SMEFT operators in Eqs.~(\ref{dim61}-\ref{dim66}) will yield FCNC processes with charged leptons at tree level.
In particular, the leptonic Kaon decays provide a complementary probe of these operators. Thus, there are constraints on the coefficients with $(\alpha,\beta)=(e,e)$ and $(\mu,\mu)$ from the Kaon decay modes $K_{L,S}\rightarrow e^+e^-, \mu^+\mu^-$. Although the matching conditions are not exactly the same for Kaon decays into neutrinos and charged leptons, we can evaluate a rough estimate on the NP scale associated with the linear combination of the coefficients responsible for both $K\to \pi \nu_e\bar{\nu_e}, \nu_\mu\bar{\nu_\mu}$ and $K_{L,S}\rightarrow e^+e^-, \mu^+\mu^-$. Under the assumption that the SM contribution has no interference with the NP contribution, there is a lower limit on the new physics scale of 83~\TeV~for $(\alpha,\beta)=(\mu,\mu)$ from $K_L\to\mu^+\mu^-$ and 20~\TeV~for $(\alpha,\beta)=(e,e)$ from $K_L\to e^+e^-$. The detailed derivation of these constraints is reported in Appendix~\ref{app2}.

More importantly, as the component with $(\alpha,\beta)=(\tau,\tau)$ does not participate in any tau lepton rare decays or leptonic charged Kaon decays at tree-level,
$K\rightarrow \pi\nu\bar\nu$ provides a unique opportunity to probe the SMEFT Wilson coefficients entering $C_{d\nu,\rm dim-6}^{V,sd\tau\tau}$.
If the NP contribution to Kaon semi-invisible decays originates only from the operator with $(\alpha,\beta)=(\tau,\tau)$, we have
\begin{align}
\mathcal{B}_{K_L\rightarrow\pi^0\nu\bar\nu}
={2 \over 3}\mathcal{B}_{K_L\rightarrow\pi^0\nu\bar\nu}^{\text{SM}}+&
4J_2^{K_L}\left({\rm Im}[C_{d\nu 1,\text{SM}}^{V,sd \tau\tau}]+{\rm Im}[C_{d\nu,\text{dim-6}}^{V,sd\tau\tau}]\right)^2,
\\
\mathcal{B}_{K^+\rightarrow\pi^+\nu\bar\nu}
={2 \over 3}\mathcal{B}_{K^+\rightarrow\pi^+\nu\bar\nu}^{\text{SM}}+&
J_2^{K^+}\Big[ \left({\rm Re}[C_{d\nu 1,\text{SM}}^{V,sd \tau\tau}]+{\rm Re}[C_{d\nu,\text{dim-6}}^{V,sd\tau\tau}]\right)^2\\&
+\left({\rm Im}[C_{d\nu 1,\text{SM}}^{V,sd \tau\tau}]+{\rm Im}[C_{d\nu,\text{dim-6}}^{V,sd\tau\tau}]\right)^2\Big]\;.
\end{align}
The first term describes the SM
contribution from decays to electron and muon neutrinos and the second term
describes the decay to tau neutrinos and receives contributions from both the SM and NP.
The LFV contributions are neglected as stated above.

\begin{figure}
\centering
\includegraphics[width=8cm]{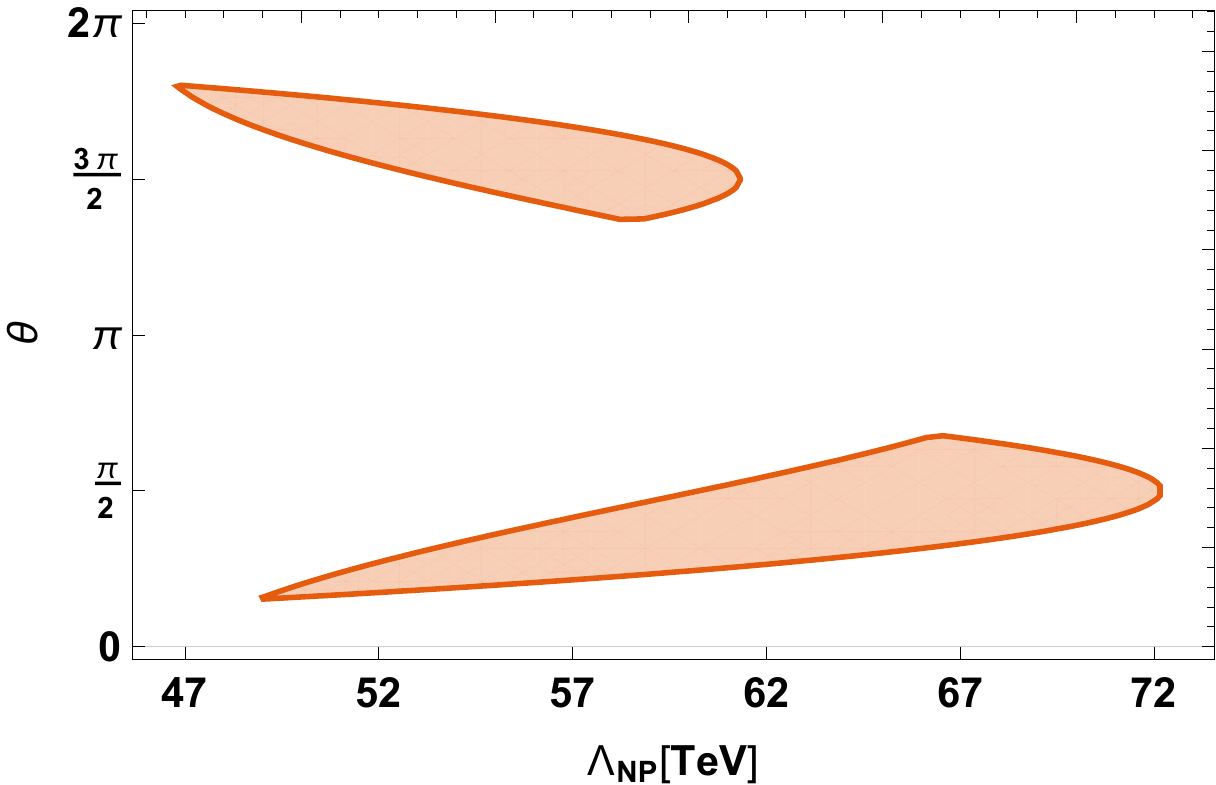}
\includegraphics[width=8cm]{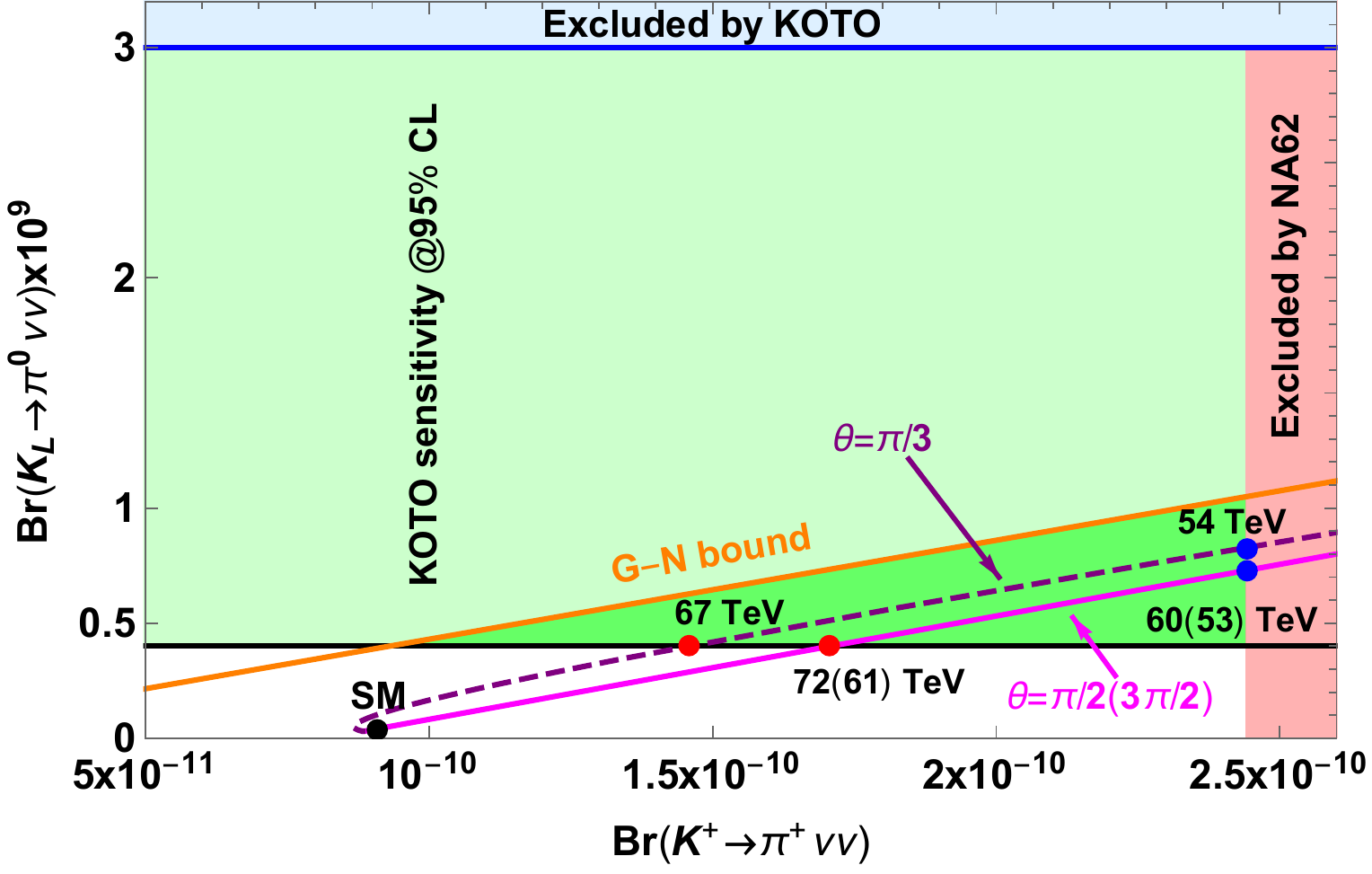}
\caption{Left: The allowed region in the $[\Lambda_{\text{NP}},\theta]$ plane. Right: The correlation of two Kaon decay branching ratios and constraint on the NP scale for the LNC operators, with $\theta=\pi/2$ or $3\pi/2$ (pink solid line) and $\theta=\pi/3$ (purple dashed line). The blue (red) points represent the lower (upper) limits of the NP scale. The sensitive region from KOTO 2016-2018 is shown in light green, and the excluded region by NA62 is in salmon. The light blue region is excluded by KOTO 2015. The black point corresponds the SM prediction.
}
\label{fig33}
\end{figure}

We further require the above results fall within the KOTO and NA62 sensitivity in
Eqs.~(\ref{koto},\ref{na62})
\begin{eqnarray}
&&3.85\times10^{-10}G_F^2<\left({\rm Im}[C_{d\nu 1,\text{SM}}^{V,sd \tau\tau}]+{\rm Im}[C_{d\nu,\text{dim-6}}^{V,sd\tau\tau}]\right)^2< 6.255\times10^{-9}G_F^2
\\
&&\left({\rm Re}[C_{d\nu 1,\text{SM}}^{V,sd \tau\tau}]+{\rm Re}[C_{d\nu,\text{dim-6}}^{V,sd\tau\tau}]\right)^2
+\left({\rm Im}[C_{d\nu 1,\text{SM}}^{V,sd \tau\tau}]+{\rm Im}[C_{d\nu,\text{dim-6}}^{V,sd\tau\tau}]\right)^2< 8.33\times10^{-10}G_F^2 \; .
\end{eqnarray}
If we denote the Wilson coefficient as
\begin{eqnarray}
C_{d\nu,\text{dim-6}}^{V,sd\tau\tau}(\Lambda_\chi)=C_{d\nu,\text{dim-6}}^{V,sd\tau\tau}(\Lambda_{\text{EW}})\equiv{e^{i\theta} \over \Lambda_{\rm NP}^2},
\end{eqnarray}
where $\theta$ denotes the phase of the Wilson coefficient.
Note that the running from the electroweak scale $\Lambda_{\text{EW}}$ to the chiral symmetry breaking scale $\Lambda_\chi$ vanishes, as the dim-6 vector operators are not renormalized at one-loop level due to QCD Ward identity. The allowed range in the plane of $[\Lambda_{\text{NP}},\theta]$ is given in the left panel of Fig.~\ref{fig33}. From this plot, we can see that the phase $\theta$ is nonzero and the NP scale is limited to
\begin{eqnarray}
\Lambda_{\rm NP} \in[47~\text{TeV},~72~\text{TeV}] \; .
\end{eqnarray}
For specific choices of $\theta=\pi/2$ or $3\pi/2$, the real part of $C_{d\nu,\text{dim-6}}^{V,sd\tau\tau}$ vanishes and both branching ratios are only governed by $\left({\rm Im}[C_{d\nu 1,\text{SM}}^{V,sd \tau\tau}]+{\rm Im}[C_{d\nu,\text{dim-6}}^{V,sd\tau\tau}]\right)^2$. Thus, in the plane of two branching ratios shown in the right panel of Fig.~\ref{fig33}, the correlation lines for these two choices coincide with each other. The NP scale resides in the range of $\Lambda_{\rm NP} \in[60~\text{TeV},~72~\text{TeV}]$ for $\theta=\pi/2$ or $\Lambda_{\rm NP} \in[53~\text{TeV},~61~\text{TeV}]$ for $\theta=3\pi/2$. We also display the case of $\theta=\pi/3$ resulting in a different line in the plane of two branching ratios.

On the other hand, for the LFV case, a similar analysis can be carried out based on Eqs.~(\ref{kotobound},~\ref{na62bound}). The SM has no interference with LFV contribution in this case. After neglecting the above LFC contribution from NP and assuming the coefficient with only one set of lepton flavors is switched on at a time, a lower limit on the NP scale associated with the LFV Wilson coefficients can be obtained as
\begin{eqnarray}
\Lambda_{\rm{NP}} = \Big|C_{d\nu,\text{dim-6}}^{V,sd\alpha\beta}\Big|^{-{1\over 2}}> 56.4 \ (63.3)~\TeV, ~\text{for}~\alpha\neq\beta \; ,
\end{eqnarray}
where the NA62 result for $K^+\to \pi^+ \nu\bar{\nu}$ at 95 (90)\% CL is taken. A stronger bound can be set if we assume the same magnitude for all LFV Wilson coefficients, that is
$C_{d\nu, {\rm dim-6}}^{V, sd e\mu}=C_{d\nu, {\rm dim-6}}^{V, sd e\tau}=C_{d\nu, {\rm dim-6}}^{V, sd \mu\tau}$
\begin{eqnarray}
\Lambda_{\rm{NP}} > 88 \ (99)~\TeV, ~\text{for}~\alpha\neq\beta \; .
\end{eqnarray}
In Refs.~\cite{Carpentier:2010ue,He:2019xxp,He:2019iqf}, there are similar analyses for LFV coefficients using the limit on $\calB(K^+\to \pi^+ \nu\bar{\nu})$ from PDG. Their limits can be translated into a bound on NP scale in our convention as 56.8 TeV in Ref.~\cite{Carpentier:2010ue} and 50 TeV in Refs.~\cite{He:2019xxp,He:2019iqf}. We can see that the new NA62 result pushes the NP scale higher. The bound obtained above is the most stringent one for the coefficients with $\tau$ flavor, compared to the bound from $\tau$ lepton LFV rare decays~\cite{He:2019iqf}. For the coefficients with $(\alpha,\beta)=(e,\mu)$ or $(\mu,e)$, the most stringent bound with $\Lambda_{\rm NP}\geq 259\,\mathrm{TeV}$ is from the charged lepton decay modes of Kaon, i.e. $K_L\to \mu^-e^+, \mu^+e^-$. See also the derivation in Appendix~\ref{app2}.

\subsection{Constraint on the LNV operator}

In this section we assume
the NP contribution from dim-6 LNC operators is negligible and
therefore only keep the SM contribution in the LNC case. Under this assumption, we focus on
the LNV NP contribution. As discussed above, the scalar LEFT
operators from one dim-7 LNV operator in the SMEFT play an important
role in the Kaon semi-invisible decays.

The Kaon invisible decays can entail constraint on the above Wilson coefficients. The effective Lagrangian for $K_L\to \nu \nu$ at the leading order is
\begin{eqnarray}\label{klinv}
\mathcal{L}_{K_L\to \nu\nu}&=& {iBF_0\over 2} \Big[ (C_{d\nu 1}^{S, sd\alpha\beta}+C_{d\nu 1}^{S, ds\alpha\beta}-C_{d\nu 2}^{S, sd\alpha\beta}-C_{d\nu 2}^{S, ds\alpha\beta})(\overline{\nu^C_\alpha}\nu_\beta)\nonumber \\
&&-(C_{d\nu 1}^{S, ds\alpha\beta\ast}+C_{d\nu 1}^{S, sd\alpha\beta\ast}-C_{d\nu 2}^{S, ds\alpha\beta\ast}-C_{d\nu 2}^{S, sd\alpha\beta\ast})(\overline{\nu_\alpha}\nu^C_\beta) \Big]K_L \; ,
\end{eqnarray}
and that for $K_S\to \nu\nu$ decay is
\begin{eqnarray}\label{ksinv}
\mathcal{L}_{K_S\to \nu\nu}&=& {iBF_0\over 2} \Big[ (C_{d\nu 1}^{S, sd\alpha\beta}-C_{d\nu 1}^{S, ds\alpha\beta}-C_{d\nu 2}^{S, sd\alpha\beta}+C_{d\nu 2}^{S, ds\alpha\beta})(\overline{\nu^C_\alpha}\nu_\beta)\nonumber \\
&&-(C_{d\nu 1}^{S, ds\alpha\beta\ast}-C_{d\nu 1}^{S, sd\alpha\beta\ast}-C_{d\nu 2}^{S, ds\alpha\beta\ast}+C_{d\nu 2}^{S, sd\alpha\beta\ast})(\overline{\nu_\alpha}\nu^C_\beta) \Big]K_S \; .
\end{eqnarray}
One can see that the processes are only induced by $|\Delta L|=2$ dim-6 operators in LEFT since they flip the helicity of one of neutrinos to allow the pseudoscalar Kaon to decay invisibly. The $|\Delta L|=0$ dim-6 operators $\calO^V_{d\nu 1/2}$ are severely suppressed by the neutrino mass because they are subject to helicity-suppression. As seen in the above subsection, only the $\calO^S_{d\nu1}$ operator is induced by one dim-7 SMEFT operator at tree-level.
By including the one-loop QCD running result for $C_{d\nu1}^{S}$ from electroweak scale $\Lambda_{\text{EW}}\approx m_W$ to the chiral symmetry breaking scale $\Lambda_\chi\approx 2~\GeV$, we obtain
\begin{eqnarray}
C_{d\nu1}^{S}(\Lambda_\chi)=1.656 \ C_{d\nu1}^{S}(\Lambda_{\text{EW}}).
\end{eqnarray}
We further assume the Wilson coefficients in Eqs.~(\ref{csds1}-\ref{cssd1}) at $\Lambda_{\text{EW}}$ scale as
\begin{eqnarray}\label{cdLQLH1}
C_{\bar{d}LQLH1}^{1\alpha 2\beta}(\Lambda_{\text{EW}})=C_{\bar{d}LQLH1}^{1\beta 2\alpha}(\Lambda_{\text{EW}})
=C_{\bar{d}LQLH1}^{2\alpha 1\beta}(\Lambda_{\text{EW}})=C_{\bar{d}LQLH1}^{2\beta 1\alpha}(\Lambda_{\text{EW}})
\equiv {1\over \Lambda_{\rm NP}^3}\delta_{\alpha\beta},
\end{eqnarray}
where $\Lambda_{\rm NP}$ denotes the NP scale above the electroweak scale.
Combining Eqs.~(\ref{klinv}-\ref{ksinv}), the matching results in Eqs.~(\ref{csds1}-\ref{cssd1}), and the naive assumption in Eq.~(\ref{cdLQLH1}), we find that there is no tree-level contribution to $K_S$ decay. For $K_L$ invisible decay, we obtain the branching ratio of invisible decay
\begin{eqnarray}
\mathcal{B}_{K_L\rightarrow \nu\nu}=2\times3\times{1\over 2}\times{m_{K_L}\over \Gamma_{K_L}^{\text{Exp}}}{1\over 16\pi}\left|0.585{BF_0v \over \Lambda_{\rm NP}^3}\right|^2\; ,
\end{eqnarray}
where the factor 2 accounts for the anti-neutrino case, the factor 3 for the 3 generations of neutrinos, the factor $1/2$ for the identical neutrinos in final states, and $1/16\pi$ for the phase space, respectively. The experimental bounds for the Kaon invisible decay were estimated in Ref.~\cite{Gninenko:2014sxa} as
\begin{eqnarray}
\mathcal{B}_{K_L\rightarrow invisible}&<&6.3\times 10^{-4} \ (95\%~ \text{C.L.}),
\\
\mathcal{B}_{K_S\rightarrow invisible}&<&1.1\times 10^{-4} \ (95\%~ \text{C.L.}).
\end{eqnarray}
The above $K_L$ bound translates into a rather weak lower limit on the new physics scale
\begin{eqnarray}
\Lambda_{\rm NP} \gtrsim 4 \ {\rm TeV} \; .
\end{eqnarray}

Given the matching results in Eqs.~(\ref{csds1}) and (\ref{cssd1}) together with the assumption in Eq.~(\ref{cdLQLH1}), the branching ratios of $K\to \pi \nu\widehat\nu$ in Eqs.~(\ref{brkl}) and (\ref{brkp}) can be simplified as
\begin{eqnarray}
\mathcal{B}_{K_L\rightarrow\pi^0\nu\widehat\nu}&=&J_1^{K_L} \sum_{\alpha\leq \beta} \left(1-{1\over2}\delta_{\alpha\beta}\right)\left|C_{d\nu 1}^{S,ds\alpha\beta}+C_{d\nu 1}^{S,sd\alpha\beta}\right|^2
+\mathcal{B}_{K_L\rightarrow\pi^0\nu\bar{\nu}}^{\text{SM}}
\nonumber\\&=&
{58.58 \over G_F^3\Lambda^6_{\rm NP}}+\mathcal{B}_{K_L\rightarrow\pi^0\nu\bar{\nu}}^{\text{SM}} \; ,
\\
\mathcal{B}_{K^+\rightarrow\pi^+\nu\widehat\nu}&=&J_1^{K^+}\sum_{\alpha\leq \beta} \left(1-{1\over2}\delta_{\alpha\beta}\right)
\left(\left|C_{d\nu 1}^{S,ds\alpha\beta}\right|^2+\left|C_{d\nu 1}^{S,sd\alpha\beta}\right|^2\right)
+\mathcal{B}_{K^+\rightarrow\pi^+\nu\bar{\nu}}^{\text{SM}} \nonumber
\\
&=&
{13\over G_F^3\Lambda^6_{\rm NP}}+\mathcal{B}_{K^+\rightarrow\pi^+\nu\bar{\nu}}^{\text{SM}} \; .
\end{eqnarray}
\begin{figure}
\centering
\includegraphics[width=10cm]{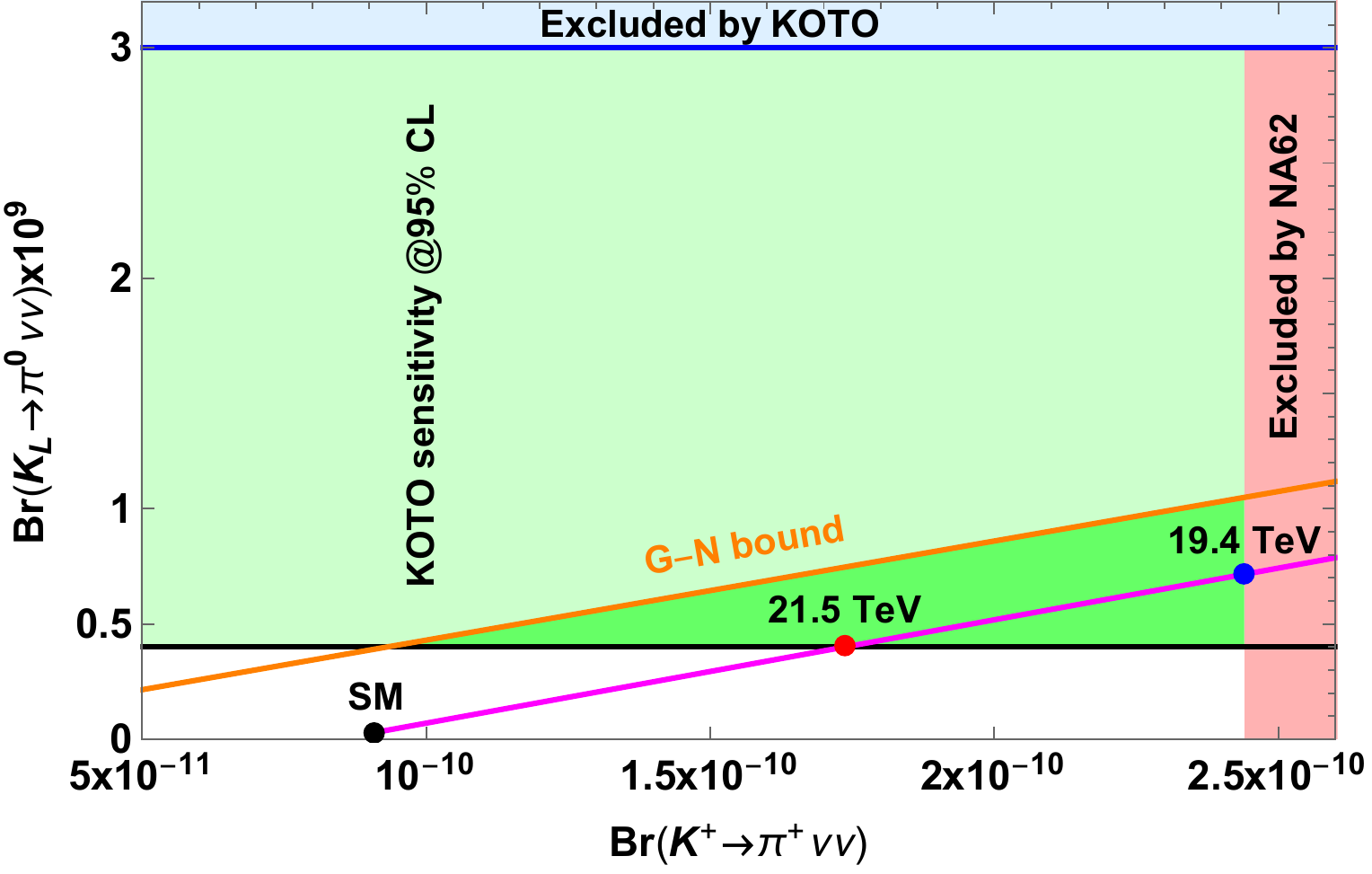}
\caption{The correlation of two Kaon decay branching ratios and constraint on the NP scale for the LNV operator. The labels and colors are the same as those in Fig.~\ref{fig33}.}
\label{fig2}
\end{figure}
There is obviously no interference between the SM contribution and the LNV contribution.
If we require those results to satisfy
the region allowed by the KOTO and NA62 upper bounds, the NP contribution resides along the pink line in Fig.~\ref{fig2} and the NP scale is highly constrained to a narrow range
\begin{eqnarray}
\Lambda_{\rm NP}\in [19.4~\TeV,~21.5~\TeV] \; .
\end{eqnarray}
See Fig.~\ref{fig2} for a more detailed illustration. This is an appealing scale both for LHC experiments and TeV scale seesaw mechanism yielding Majorana neutrino mass. This interpretation highly depends on the precision of the measurements and the simple assumption in Eq.~(\ref{cdLQLH1}). If we take the upper bound on $\mathcal{B}_{K^+\rightarrow\pi^+\nu\bar\nu}$ from the NA62 experiment at 90\% CL, that is $\mathcal{B}_{K^+\rightarrow\pi^+\nu\widehat\nu}<1.85\times 10^{-10}$~\cite{NA62}, the contribution of the LNV operator together with the assumed Wilson coefficients in Eq.~\eqref{cdLQLH1} is almost excluded. The analysis of the 2018 NA62 data would indicate if LNV operator can precipitate the discrepancy under the assumption.

The operator in Eq.~\eqref{dim7operator} can also contribute to neutrinoless double beta ($0\nu\beta\beta$) decay. The NP scale from this process is constrained to be larger than $\calO(100~\TeV)$~\cite{Cirigliano:2017djv,Liao:2019tep}. Constraints from $K\to\pi\nu\nu$ are complementary and provide a similar sensitivity, because they constrain the quark flavor violating Wilson coefficients with an $s$- and a $d$-quark and any arbitrary generations of the lepton fields in the operator after relaxing the assumption in Eq.~\eqref{cdLQLH1}.

\section{Discussions and Conclusions}
\label{sec:Con}

In the above analysis, we focus on the contact interactions from effective operators composed of $s, d$ quarks and two neutrinos for $s\to d \nu\widehat{\nu}$ transition, that is the so-called short-distance (SD) contribution. In addition, there exist the long-distance (LD) contributions to
$K\to\pi\nu\bar{\nu}$ from the heavy NP parameterized by the dim-6 LNC operators in SMEFT. The LD contributions are mediated by light
charged leptons, neutrinos or light meson propagators in the $\chi$PT picture. In the $K\to\pi\nu\bar\nu$ processes, it turns out that the LD contributions are negligible compared to the SD contribution and can be safely ignored. The detailed analysis
is included in Appendix~\ref{sec:LD}.

In summary, we investigate the implication of $K\to \pi \nu \bar{\nu}$ from the recent KOTO and NA62 measurements for generic neutrino interactions in effective field theories.
The interactions between quarks and left-handed SM neutrinos are first described by the LEFT below electroweak scale. We match them to $\chi$PT at the chiral symmetry breaking scale to calculate the branching fractions of Kaon semi-invisible decays and match them up to the SMEFT to constrain new physics above the electroweak scale.

In the framework of effective field theories, we prove that the Grossman-Nir bound is valid for both dim-6 and dim-7 LEFT operators in $\chi$PT, and the dim-6 scalar and vector operators dominantly contribute to the $K\to \pi \nu \widehat{\nu}$ transition. They are induced by multiple dim-6 LNC operators and one dim-7 LNV operator in the SMEFT, respectively. After providing a generic constraint on the relevant Wilson coefficients, we separately discuss the constraints on the two kinds of operators. The LNC vector operators lead to interference with the SM contribution. We consider the lepton-flavor-conserving case and evaluate the constraints from Kaon leptonic decay modes $K_L\to e^+e^-, \mu^+\mu^-$. For the $\tau\tau$ component in the $s\to d$ transition, the $K\to \pi \nu \bar{\nu}$ decays provide the only sensitive probe. We find the NP scale associated with the $\tau\tau$ Wilson coefficient is limited to be $\Lambda_{\rm NP} \in[47~\text{TeV},~72~\text{TeV}]$.

One the other hand,
we assume the NP
contribution from dim-6 LNC operators is negligible and therefore
only keep the SM prediction in the LNC case. As a result, the scalar LEFT
operators from one dim-7 LNV operator in the SMEFT dominates the
$K\to \pi \nu\widehat\nu$ decay. We find that the $K_L$ invisible decay $K_L\to \nu\nu$
places a
weak bound on the new physics scale for the
LNV operator.
As there is no interference with the SM
contribution, the constraint on the NP scale from $K\to \pi \nu\bar\nu$ is rather precise and
resides in a narrow range $\Lambda_{\rm NP}\in
[19.4~\TeV,~21.5~\TeV]$.

\appendix

\section{The amplitudes and partial widths of $K\to \pi\nu\widehat{\nu}$}

\label{sec:Kpinunu}
In this appendix we present details of the calculation of $K\to \pi \nu \widehat\nu$. For the process of $K(p_K)\to \pi(p_\pi)+\overline{\nu_\alpha}(p_1){\overline{\nu}_\beta}(p_2)/\nu_\alpha(p_1)\nu_\beta(p_2)/\nu_\alpha(p_1){\overline{\nu}_\beta}(p_2)$, the amplitudes for $K_L$ decay are
\begin{eqnarray}
iM_1^{\Delta L=-2}(K_L)&=&{iB\over 4}
2\left(C_{d\nu1}^{S,sd\alpha\beta}+C_{d\nu2}^{S,sd\alpha\beta}
+C_{d\nu1}^{S,ds\alpha\beta}+C_{d\nu2}^{S,ds\alpha\beta}\right)\nu^T(p_1) C \nu(p_2)\; , \nonumber \\
iM_1^{\Delta L=2}(K_L)&=&{iB\over 4}
2\left(C_{d\nu1}^{S,ds\alpha\beta*}+C_{d\nu2}^{S,ds\alpha\beta*}
+C_{d\nu1}^{S,sd\alpha\beta*}+C_{d\nu2}^{S,sd\alpha\beta*}\right)\overline{\nu}(p_1)C \overline{\nu^T}(p_2)\; , \nonumber \\
iM_2^{\Delta L=0}(K_L)&=&{i\over 4}
\left(C_{d\nu 1}^{V,sd\alpha\beta}+C_{d\nu 2}^{V,sd\alpha\beta}-C_{d\nu1}^{V,ds\alpha\beta}-C_{d\nu 2}^{V,ds\alpha\beta}\right) (p_K+p_\pi)_\mu \overline{\nu}(p_1) \gamma^\mu \nu(p_2)\; ,
\end{eqnarray}
and those for $K^+$ decay are
\begin{eqnarray}
iM_1^{\Delta L=-2}(K^+)&=&{-iB\over 2}
2
\left(C_{d\nu1}^{S,sd\alpha\beta}+C_{d\nu2}^{S,sd\alpha\beta}\right)\nu^T(p_1) C \nu(p_2)\; , \nonumber \\
iM_1^{\Delta L=2}(K^+)&=&{-iB\over 2}
2\left(C_{d\nu1}^{S,ds\alpha\beta*}+C_{d\nu2}^{S,ds\alpha\beta*}\right)\overline{\nu}(p_1) C\overline{\nu^T}(p_2)\; , \nonumber \\
\label{kps}
iM_2^{\Delta L=0}(K^+)&=&{-i\over 2}
\left(C_{d\nu 1}^{V,sd\alpha\beta}+C_{d\nu 2}^{V,sd\alpha\beta}\right) (p_K+p_\pi)_\mu \overline{\nu}(p_1) \gamma^\mu \nu(p_2)\; ,
\end{eqnarray}
with $\alpha,\beta=e,\mu,\tau$. The factor of $2$ comes from the symmetry property of the operators and the corresponding Wilson coefficients for $|\Delta L|=2$.
We neglect the masses of final state neutrinos and thus the different amplitudes do not interfere. After summing over all possible flavors, the flavor- and spin- summed squared matrix elements read
\begin{eqnarray}
\sum_{all}|M(K_L)|^2&=&\sum_{\alpha\leq \beta}\sum_{\rm spin}\left( |M_1^{\Delta L=-2}(K_L)|^2+|M_1^{\Delta L=2}(K_L)|^2\right)+\sum_{\alpha, \beta}\sum_{\rm spin}|M_2^{\Delta L=0}(K_L)|^2 \nonumber
\\
&=&2\sum_{\alpha\leq \beta} \left(1-{1\over2}\delta_{\alpha\beta}\right)\frac{B^2}{4}\left|C_{d\nu1}^{S,sd\alpha\beta}+C_{d\nu2}^{S,sd\alpha\beta}
+C_{d\nu1}^{S,ds\alpha\beta}+C_{d\nu2}^{S,ds\alpha\beta}\right|^2s \nonumber
\\
&+&\frac{1}{4}\sum_{\alpha, \beta}\left|C_{d\nu 1}^{V,sd\alpha\beta}+C_{d\nu 2}^{V,sd\alpha\beta}-C_{d\nu1}^{V,ds\alpha\beta}-C_{d\nu 2}^{V,ds\alpha\beta}\right|^2  \left((m_K^2-t)(t-m_\pi^2)-st\right) \nonumber
\\ \\
\sum_{all}|M(K^+)|^2&=&\sum_{\alpha\leq \beta}\sum_{\rm spin}\left( |M_1^{\Delta L=-2}(K^+)|^2+|M_1^{\Delta L=2}(K^+)|^2\right)+\sum_{\alpha, \beta}\sum_{\rm spin}|M_2^{\Delta L=0}(K^+)|^2 \nonumber
\\
&=&\sum_{\alpha\leq \beta} \left(1-{1\over2}\delta_{\alpha\beta}\right)B^2\left(\left|C_{d\nu1}^{S,sd\alpha\beta}+C_{d\nu2}^{S,sd\alpha\beta}\right|^2+\left|C_{d\nu1}^{S,ds\alpha\beta}+C_{d\nu2}^{S,ds\alpha\beta}\right|^2
\right)s \nonumber
\\
&+&\sum_{\alpha, \beta}\left|C_{d\nu 1}^{V,sd\alpha\beta}+C_{d\nu 2}^{V,sd\alpha\beta}\right|^2\left((m_K^2-t)(t-m_\pi^2)-st\right),
\end{eqnarray}
where $s=(p_1+p_2)^2$ and $t=(p_2+p_\pi)^2$. Here $\alpha\leq\beta$ means that for $|\Delta L|=2$ we take $\alpha\beta=ee,e\mu,e\tau,\mu\mu,\mu\tau,\tau\tau$ flavor configurations. One should note that the flavor indices $\alpha,\beta$ are implicitly summed over in the Lagrangian, while in the above amplitudes they denote the specific neutrino flavors in final states.
The overall factor 2 in the second line for $K_L$ decay is because the contributions from $M_1^{\Delta L=2}(K_L)$ and $M_1^{\Delta L=-2}(K_L)$ are the same for any pair of  $(\alpha,\beta)$. The $-1/2\delta_{\alpha\beta}$ accounts for the double counting of final state phase space for identical particles.

The partial decay width can be expressed as
\begin{eqnarray}
\Gamma_{K\rightarrow \pi\nu\widehat\nu}=\frac{1}{2m_K}\frac{1}{128\pi^3m_K^2}\int_0^{(m_K-m_\pi)^2} ds \int dt |\mathcal{M}(K\rightarrow \pi \nu\nu)|^2 \; ,
\label{decaywidth}
\end{eqnarray}
where the integration interval of $t$ is
\begin{eqnarray}
t\in \left[\left(E_2^*+E_3^*\right)^2-\left(E_2^{*}+\sqrt{E_3^{*2}-m_\pi^2}\right)^2,~\left(E_2^*+E_3^*\right)^2-\left(E_2^{*}-\sqrt{E_3^{*2}-m_\pi^2}\right)^2\right],
\end{eqnarray}
with
\begin{eqnarray}
E_2^*=\frac{1}{2}\sqrt{s},~E_3^*=\frac{1}{2}\frac{m_K^2-m_\pi^2-s}{\sqrt{s}} \; .
\end{eqnarray}
The partial widths of Kaon semi-invisible decays are obtained by performing the $t$ integral
\begin{eqnarray}\nonumber
{d\Gamma_{K_L\rightarrow \pi^0\nu\widehat\nu}\over d s}&=&
{B^2\over 2^9\pi^3m_K^3}
\sum_{\alpha\leq \beta} \left(1-{1\over2}\delta_{\alpha\beta}\right)\left|C_{d\nu1}^{S,sd\alpha\beta}+C_{d\nu2}^{S,sd\alpha\beta}
+C_{d\nu1}^{S,ds\alpha\beta}+C_{d\nu2}^{S,ds\alpha\beta}\right|^2
\\\nonumber
&&\times s\left((m_K^2+m_\pi^2-s)^2-4m_K^2m_\pi^2\right)^{1/2}
\\\nonumber
&&
+{1\over 3\cdot2^{11}\pi^3m_K^3}\sum_{\alpha, \beta}\left|C_{d\nu 1}^{V,sd\alpha\beta}+C_{d\nu 2}^{V,sd\alpha\beta}-C_{d\nu1}^{V,ds\alpha\beta}-C_{d\nu 2}^{V,ds\alpha\beta}\right|^2
\\
&&\times \left((m_K^2+m_\pi^2-s)^2-4m_K^2m_\pi^2\right)^{3/2},
\label{PW1}
\\\nonumber
{d\Gamma_{K^+\rightarrow \pi^+\nu\widehat\nu}\over d s}&=&
{B^2\over 2^8\pi^3m_K^3}\sum_{\alpha\leq \beta} \left(1-{1\over2}\delta_{\alpha\beta}\right)\left(\left|C_{d\nu1}^{S,sd\alpha\beta}+C_{d\nu2}^{S,sd\alpha\beta}\right|^2+\left|C_{d\nu1}^{S,ds\alpha\beta}+C_{d\nu2}^{S,ds\alpha\beta}\right|^2
\right)
\\\nonumber
&&\times s\left((m_K^2+m_\pi^2-s)^2-4m_K^2m_\pi^2\right)^{1/2}
\\
&&
+{1\over 3\cdot2^{9}\pi^3m_K^3}\sum_{\alpha, \beta}\left|C_{d\nu 1}^{V,sd\alpha\beta}+C_{d\nu 2}^{V,sd\alpha\beta}\right|^2
\left((m_K^2+m_\pi^2-s)^2-4m_K^2m_\pi^2\right)^{3/2}.
\label{PW2}
\end{eqnarray}

\section{The constraints from leptonic Kaon decays}
\label{app2}

The current experimental constraints on the branching ratios of $K_S\to e^+e^-$~\cite{Ambrosino:2008zi} and $K_S\to \mu^+\mu^-$~\cite{LHCb:2019aoh} are
\begin{align}\label{KSllexp}
	\calB(K_S\to e^+ e^-) & < 9\times 10^{-9}\; , &
	\calB(K_S\to \mu^+ \mu^-) &< 2.4\times 10^{-10}\; , &
\end{align}
at 90\% CL. The lepton-flavor-conserving modes of $K_L$ decays have been measured~\cite{Tanabashi:2018oca}
\begin{align}\label{KLllexp}
	\calB(K_L\to e^+ e^-) & = 9^{+6}_{-4} \times 10^{-12}\; , &
	\calB(K_L\to \mu^+ \mu^-) & = (6.84\pm0.11) \times 10^{-9}\;.
\end{align}
Their SM predictions are~\cite{Valencia:1997xe,GomezDumm:1998gw,DAmbrosio:2017klp}
\begin{align}\label{KLllSM}
\calB(K_L\to e^+ e^-)_{\rm SM} & \approx 9 \times 10^{-12}\; , &
\calB(K_L\to \mu^+ \mu^-)_{\rm SM} & = (6.85\pm0.86) \times 10^{-9}\;.
\end{align}
We match the SMEFT Wilson coefficients in Eq.~\eqref{dim61}, relevant for Kaon physics, to the four vector operators in LEFT
\begin{align}
\calO^{V}_{de1}&=(\overline{d_L}\gamma_\mu d_L)(\overline{e_L}\gamma_\mu e_L),
	       &
\calO^{V}_{de2}&=(\overline{d_L}\gamma_\mu d_L)(\overline{e_R}\gamma_\mu e_R),
	       \\
\calO^{V}_{de3}&=(\overline{d_R}\gamma_\mu d_R)(\overline{e_L}\gamma_\mu e_L),
	       &
\calO^{V}_{de4}&=(\overline{d_R}\gamma_\mu d_R)(\overline{e_R}\gamma_\mu e_R).
\end{align}
The matching condition is given by\footnote{We do not include tensor operators $\calO_{dW}$ and $\calO_{dB}$ which are suppressed in $\chi$PT power counting relative to the vector-type operators, and also the scalar operators $\calO_{dH}$ which are suppressed by the SM Yukawa couplings.}
\begin{eqnarray}\label{cde1}
C^{V,sd\alpha\beta}_{de1}&=&
D_{xs}^*D_{yd}\left(C_{l q}^{(1),\alpha\beta xy}+C_{l q}^{(3),\alpha\beta xy}-(1-2s_W^2)\left( C_{Hq}^{(1),xy}+C_{Hq}^{(3),xy}\right)\delta_{\alpha\beta}\right),
\\
C^{V,sd\alpha\beta}_{de2}&=&D_{xs}^*D_{yd}2s_W^2\left(C_{Hq}^{(1),xy}+C_{Hq}^{(3),xy}\right)\delta_{\alpha\beta},
\\
C^{V,sd\alpha\beta}_{de3}&=&C_{l q}^{\alpha\beta 21}-(1-2s_W^2)C_{Hd}^{sd}\delta_{\alpha\beta},
\\\label{cde4}
C^{V,sd\alpha\beta}_{de4}&=&2s_W^2C_{Hd}^{sd}\delta_{\alpha\beta} \; .
\end{eqnarray}
The lepton-flavor-conserving decay widths are
\begin{align}\label{KXll}
\calB(K_X\to\ell^+\ell^-) &=\calB(K_X\to\ell^+\ell^-)_{\rm SM}+ \frac{F_K^2 m_\ell^2}{32\pi \Gamma_{K_X}}\sqrt{m_{K_X}^2-4m_\ell^2}\Big|C_{K_X,\text{dim-6}}^{V,sd\ell\ell}\Big|^2+\fbox{I.T.},
\end{align}
where $X=S,L$ and $F_K$ is the physical Kaon decay constant. \fbox{I.T.} stands for the interference term between the SM part and the NP part.
Here the \fbox{I.T.} contribution is larger than the pure NP squared term in the above equation. To make a conservative estimation of the NP scale, we simply neglect the interference term below. Note that, once including the additional interference term, we should obtain a more stringent limit on the NP scale for $ee$ and $\mu\mu$ coefficients. The NP contribution is the linear combination of Wilson coefficients in Eqs.~(\ref{cde1}-\ref{cde4}) and takes the form as
\begin{align}
C_{K_S,\text{dim-6}}^{V,sd\ell\ell}&=\left(C_{de1}^{V,sd\ell\ell}-C_{de2}^{V,sd\ell\ell} - C_{de3}^{V,sd\ell\ell}+ C_{de4}^{V,sd\ell\ell}\right)-c.c. \; ,
\\
C_{K_L,\text{dim-6}}^{V,sd\ell\ell}&=\left(C_{de1}^{V,sd\ell\ell} +C_{de2}^{V,sd\ell\ell} + C_{de3}^{V,sd\ell\ell}+ C_{de4}^{V,sd\ell\ell}\right)+c.c. \; .
\end{align}
Considering the physical lifetime of $K_S$ and $K_L$ and the experimental constraints in Eqs.~(\ref{KSllexp},~\ref{KLllexp}), the stronger limit on NP scale is set by the $K_L$ decays. After neglecting the interference term in Eq.~\eqref{KXll} and subtracting the SM contribution in Eq.~\eqref{KLllSM}, the NP scales associated with the $ee$ and $\mu\mu$ coefficients are constrained to be
\begin{align}
\Lambda_{\rm NP}= \Big|C_{K_L,\text{dim-6}}^{V,sd\mu\mu}\Big|^{- {1\over 2}}\geq 83~\TeV,
 \\
\Lambda_{\rm NP}= \Big|C_{K_L,\text{dim-6}}^{V,sdee}\Big|^{- {1\over 2}}\geq 20~\TeV\;.
\end{align}
Finally, we quote the result for the LFV mode $K_L\to \mu^\pm e^\mp$
\begin{align}
	\calB(K_L\to \mu^+ e^-)=& \frac{(m_{K_L}^2-m_\mu^2)^2 F_K^2m_\ell^2}{64\pi m_{K_L}^3\Gamma_{K_L}} \Big|C_{K_L,\rm LFV}^{V,sd e\mu}\Big|^2 \; ,\\
	\calB(K_L\to \mu^- e^+)=& \frac{(m_{K_L}^2-m_\mu^2)^2 F_K^2m_\ell^2}{64\pi m_{K_L}^3\Gamma_{K_L}} \Big|C_{K_L,\rm LFV}^{V,sd \mu e}\Big|^2 \; ,
\end{align}
\begin{equation}
	\Big|C_{K_L,\rm LFV}^{V,sd \ell\ell^\prime}\Big|^2=\Big| C_{de2}^{V,sd\ell\ell^\prime}  -C_{de4}^{V,sd\ell\ell^\prime} -c.c.\Big|^2
		 + \Big| C_{de1}^{V,sd\ell\ell^\prime} -C_{de3}^{V,sd\ell\ell^\prime} -c.c.\Big|^2 \; .
	 \end{equation}
The upper limit on the lepton-flavor-violating decay of
$
	\calB(K_L\to e^\pm \mu^\mp) < 0.47 \times 10^{-11}
$
at 90\% CL~\cite{Tanabashi:2018oca} leads to a constraint on the NP scale of
\begin{equation}
	\Lambda_{\rm NP} = \Big|C_{K_L,\rm LFV}^{V,sd \ell\ell^\prime}\Big|^{-\tfrac12}  \geq\, 259 \ \mathrm{TeV} \; ,
\end{equation}
with $(\ell,\ell^\prime)=(e,\mu)$ or $(\mu,e)$.

\section{Long-distance contributions from dim-6 operators}
\label{sec:LD}

\begin{figure}
\centering
\includegraphics[width=15cm]{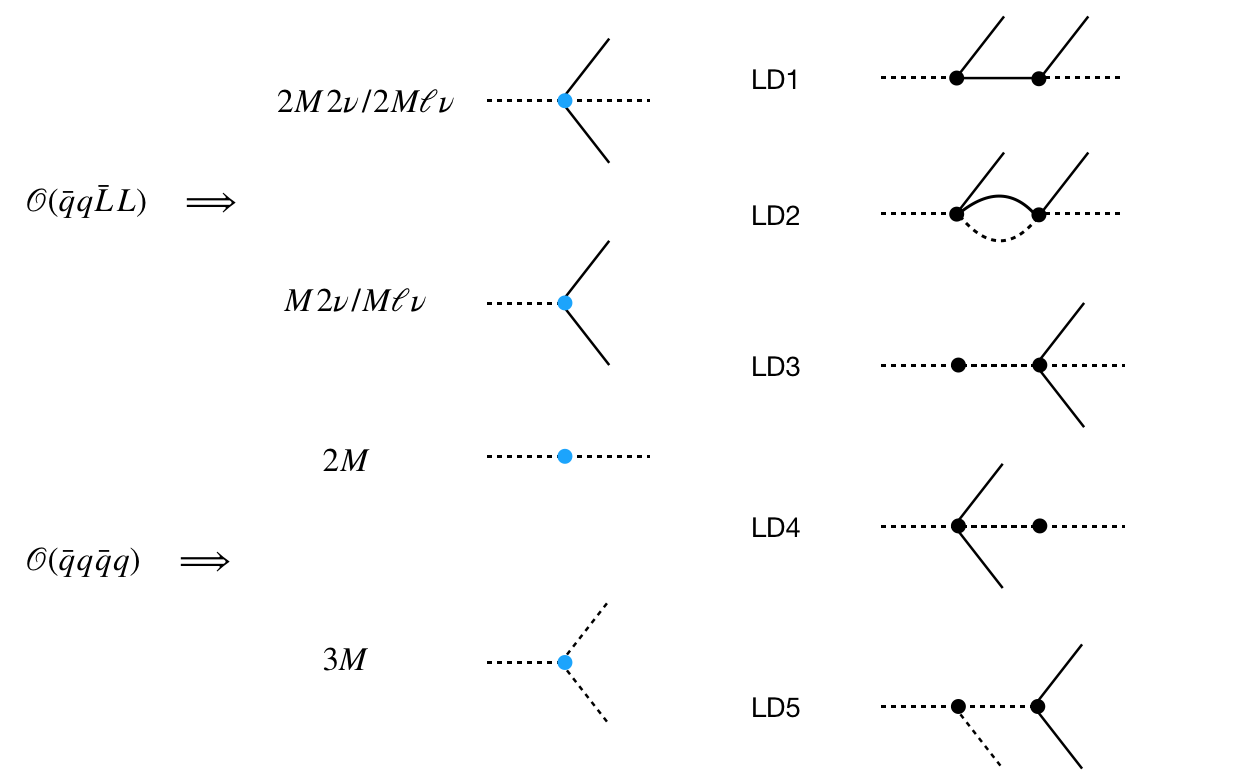}
\caption{The topologies of LD contribution to $K\rightarrow\pi \nu\bar{\nu}$ in the context of $\chi$PT with the light charged leptons, neutrinos and Goldstone mesons as dynamical degrees of freedom. There also exist other topologies like the pure meson loops which we do not show here, since they are severely suppressed by additional loop factors and a $\chi$PT factor $p/\Lambda_\chi$ relative to the shown diagrams.}
\label{Fig4}
\end{figure}

In this Appendix, we estimate the long distance (LD) contributions to
$K\to\pi\nu\bar{\nu}$ from the heavy NP parameterized by the dim-6
LNC operators in SMEFT. The LD contributions are mediated by light
charged leptons, neutrinos or light meson propagators in the
$\chi$PT picture. In Fig.~\ref{Fig4} we categorize the possible topologies
for the LD contribution from the dim-6 two-quark-two-lepton
operators $\calO(\bar{q}q\bar{L}L)$ and four-quark operators
$\calO(\bar{q}q\bar{q}q)$ in SMEFT. The dashed and solid lines
represent the possible meson and lepton fields, respectively.

The LD contributions mediated by neutrinos are suppressed compared to the SD contribution. In the Feynman diagrams for this kind of LD contribution, the vertex connecting the Kaon state involves the same Wilson coefficients as the SD case and the other vertex leads to one additional suppression factor $G_F$. Hence, we find that they are suppressed by $F_0^2G_F\sim 10^{-7}$.	

The LD contributions mediated by charged leptons can be induced by charged-current vector and/or scalar operators. The contribution from scalar operators is strongly constrained by charged pseudoscalar meson decays. The branching ratio for $M^+=K^+,\pi^+$ is
\begin{align}
	\calB(M^+ \to \ell^+ \nu) & \propto
	\sum_i \Big\{
		m_\ell^2\, \left|C_{du\nu\ell}^{V,\alpha\beta i\ell}\right|^2
		+ B^2\, \left|C_{du\nu\ell}^{S,\alpha\beta i \ell}\right|^2
	+\fbox{I.T.},
	\Big\}\;,
\end{align}
where we sum over neutrino flavor $i$ and $C_{du\nu\ell}^{S,\alpha\beta i\ell}\equiv C_{du\nu\ell1}^{S,\alpha\beta i \ell} - C_{du\nu\ell2}^{S,\alpha\beta i\ell}$ ($C_{du\nu\ell}^{V,\alpha\beta i\ell}\equiv C_{du\nu\ell1}^{V,\alpha\beta i\ell}-C_{du\nu\ell2}^{V,\alpha\beta i\ell}$) denotes the scalar (vector) operator contribution. The SM contribution is helicity-suppressed and given by $C_{du\nu\ell1}^{V,\alpha\beta i\ell}=2\sqrt{2}G_F V_{\alpha\beta}$, while NP scalar contributions do not suffer from helicity-suppression. Requiring that the NP scalar contribution can be at most as large as the SM contribution translates into $|C_{du\nu\ell}^{S,\alpha\beta i\ell}|\lesssim 2 \sqrt{2}G_F |V_{\alpha\beta}| m_\ell/B$. The same scalar contribution enters the LD contribution to $K\to\pi\nu\bar\nu$ mediated by a charged lepton.
After considering the phase space, the squared matrix element
$
\left|\mathcal{M}_{K^+\rightarrow\pi^+\nu\bar{\nu}}^{\Delta L=0}\right|^2
=\sum_{\alpha, \beta} |\hat{C}_{\alpha\beta}|^2\left((m_K^2-t)(t-m_\pi^2)-st\right)
$
is determined in terms of
\begin{eqnarray}
|\hat{C}_{\alpha\beta}|^2&=&\left|C_{d\nu 1}^{V,sd\alpha\beta}+C_{d\nu 2}^{V,sd\alpha\beta}\right|^2
+
3\times10^{-3} \left| B^2 C_{du\nu e}^{S,su\alpha\mu} C_{du\nu e}^{S,du\mu\beta*}\right|^2+\fbox{I.T.}\;.
\end{eqnarray}
The second term is suppressed by a factor of $\calO(10^{-5})$ and consequently the LD scalar contribution is sub-dominant.

In the SM, the LD contribution induced by vector operators is suppressed by $\calO(10^{-4})$ relative to the SD contribution~\cite{Hagelin:1989wt}. As the NP contribution to charged current operators is at most of the same order as the SM contribution, the LD contribution from vector operators is negligible.

Similarly, the meson-mediated tree-level contributions (LD3 and LD4) are suppressed by $\calO(10^{-4})$ with respect to the SD contribution~\cite{Hagelin:1989wt,Lu:1994ww} in the SM, while the one-loop contribution LD2 is of the same order as the LD1 contribution in the SM~\cite{Rein:1989tr,Buchalla:1998ux}. To our knowledge, there is no general LEFT analysis of LD contributions to $K\to \pi \nu\bar\nu$. As four quark operators with $\Delta S=1$ directly contribute to hadronic Kaon decays of which many have been measured at sub-percent level precision, we expect that similar conclusions hold for NP contributions mediated by meson exchange.
In summary, currently it is safe to neglect long-distance contributions to $K\to\pi\nu\bar\nu$.
	
\acknowledgments
We would like to thank Xiao-Gang He, Yi Liao, Jusak Tandean and Jian Zhang for very useful discussions and communication.
TL is supported by the National Natural Science Foundation of China (Grant No. 11975129) and ``the Fundamental Research Funds for the Central Universities'', Nankai University (Grants No. 63191522, 63196013). XDM is supported by the MOST (Grant No. MOST 106-2112-M-002-003-MY3).

\bibliography{refs}

\end{document}